\documentclass{svjour3}                     
\smartqed  

\usepackage{color}
\usepackage{graphicx}
\usepackage{bm}

\newcommand\Rey{\mbox{\textit{Re}}}  
\newcommand\Fr{\mbox{\textit{Fr}}}  

\newcommand{\dt}{\Delta t}

\newcommand{\nb}{\nabla_{\bot}}

\newcommand\etc{etc.\ }
\newcommand\eg{e.g.\ }

\newcommand\B{{\textsf B}}
\newcommand\E{{\textsf E}}
\newcommand\A{{\textsf A}}

\begin{document}

\title{Shallow water modeling of rolling pad instability in liquid metal batteries}

\author{Oleg Zikanov} 
\institute{Department of Mechanical Engineering, University of Michigan-Dearborn, 48128-1491 MI, USA\\
\email{zikanov@umich.edu}}

\date{\today}

\maketitle

\begin{abstract}
Magnetohydrodynamically induced interface instability in liquid metal batteries is analyzed. The batteries are represented by a simplified system in the form of a rectangular cell, in which strong vertical electric current flows through  three horizontal layers: the layer of a heavy metal at the bottom, the layer of a light metal at the top, and the layer of electrolyte in the middle. A new two-dimensional nonlinear model based on the {conservative shallow water approximation} is derived and utilized in a numerical study. It is found that in the case of small density difference between the electrolyte and one of the metals, the instability closely resembles the rolling pad instability observed earlier in the aluminum reduction cells. When the two electrolyte-metal density differences are comparable, the  dynamics of unstable systems is more complex and characterized by interaction between two nearly symmetric or antisymmetric interfacial waves.
\end{abstract}


\maketitle

\section{Introduction}\label{sec:intro}

The liquid metal battery is a promising conceptual device for stationary energy storage \cite{Kim:2013}. Recently developed small-scale (about few cm)  laboratory prototypes have demonstrated technical feasibility of the concept and its advantages of higher efficiency and longer operational life in comparison to traditional solid-electrode batteries (see \eg \cite{Bradwell:2012,Kim:2013calcium,Wang:2014,Ouchi:2016,Xu:2016}). Scaling up the concept to large commercially attractive devices has not, however, yet been achieved.  As we discuss below, the hydrodynamic effects, in particular that of electromagnetically modified interfacial waves, are expected to be non-negligible factors of this transition. 

\begin{figure}
\begin{center}
\includegraphics[width=0.65\textwidth]{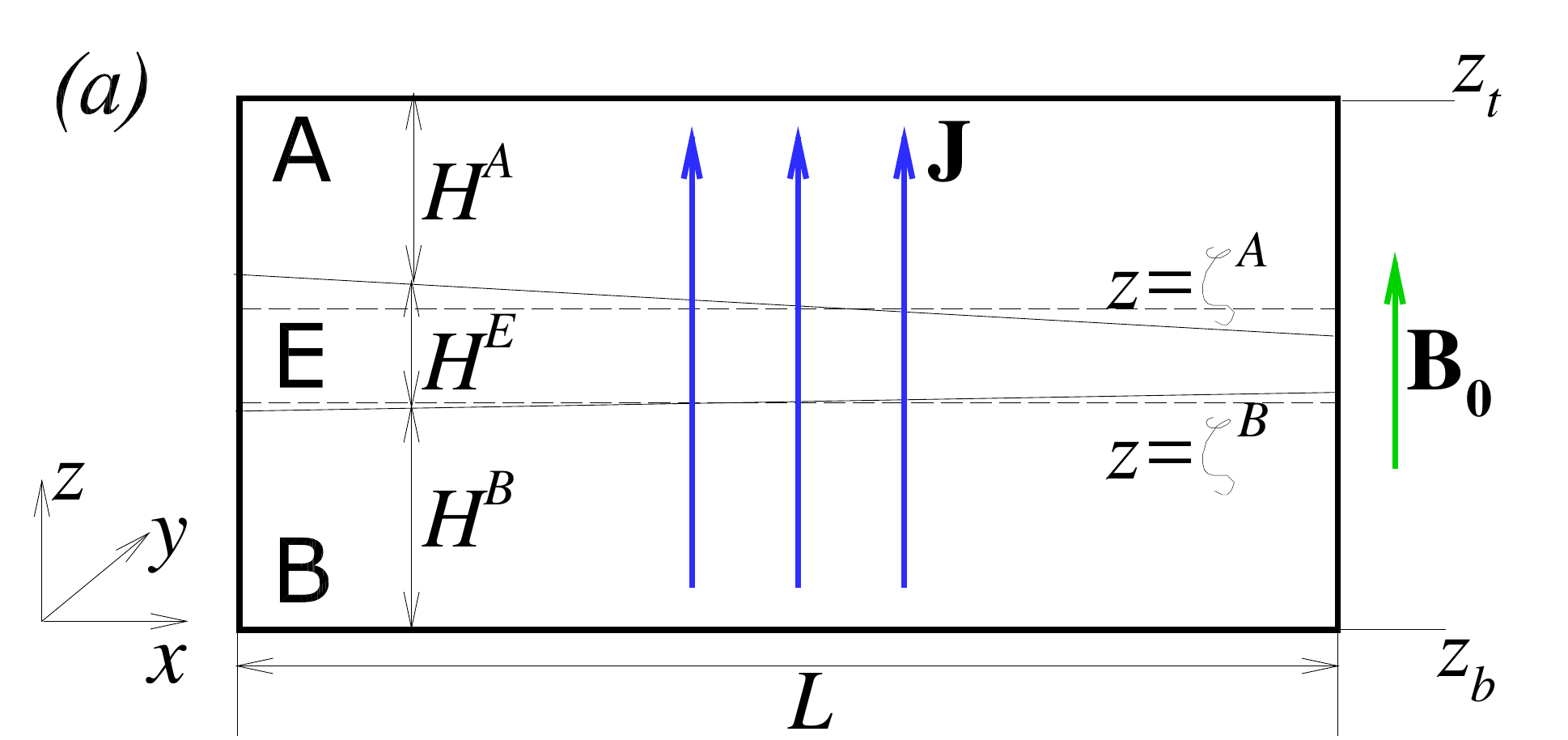}\\
\includegraphics[width=0.65\textwidth]{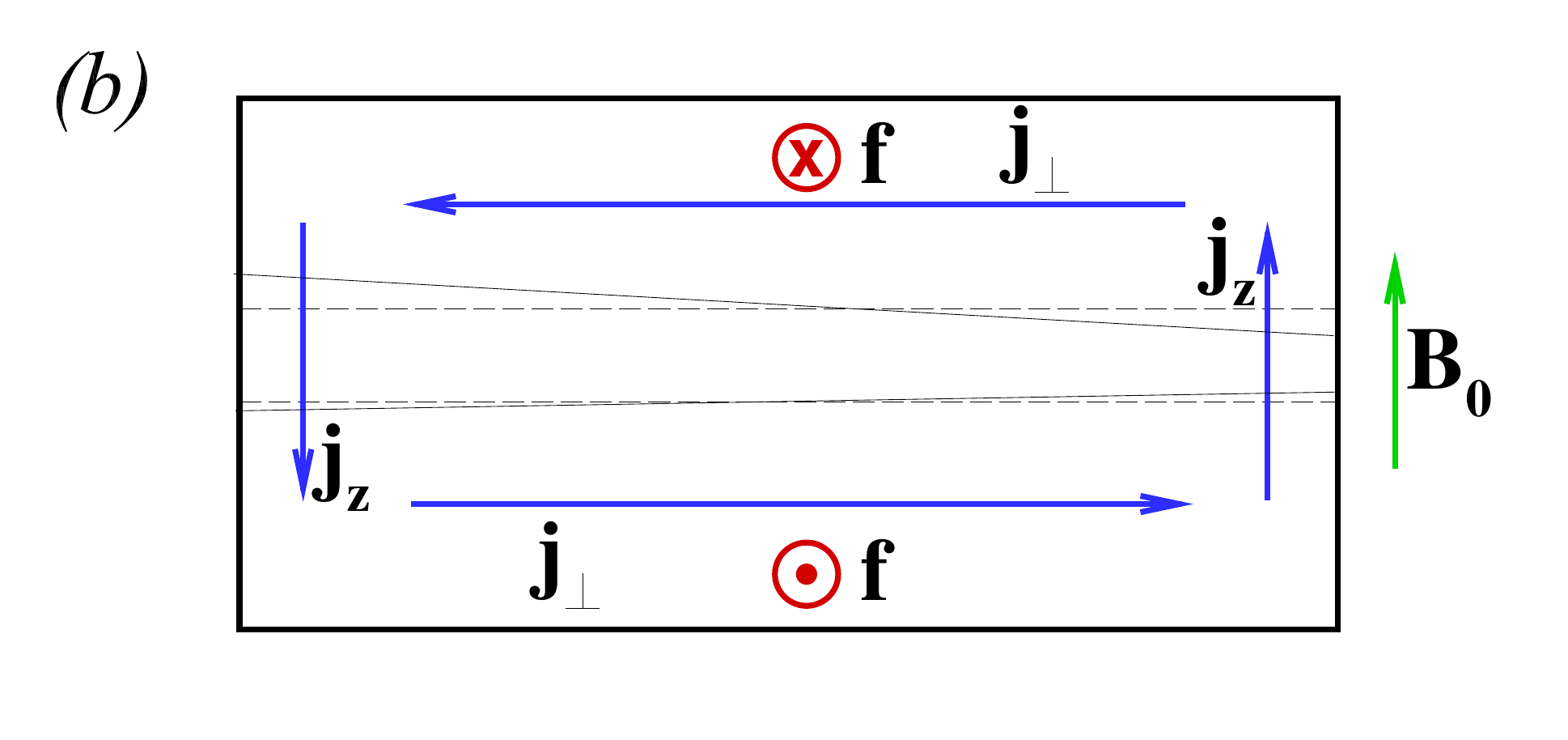}\\
\caption{\emph{(a}),  Model of a liquid metal battery considered in the study. The parameters and variables shown in the picture are introduced in section \ref{sec:modmeth}. \emph{(b}), Schematic representation of the physical mechanism of the rolling pad instability discussed later in the text.}
\label{fig1}
\end{center}
\end{figure}

In this paper we consider one version of the battery, which, on the level of simplification sufficient for out purposes, can be viewed as the system illustrated in Fig.~\ref{fig1}a. It is a cuboid or cylindrical cell filled with three liquid layers: the layer of a heavy metal (\eg Bi, Sb, Zn, or PbSb) at the bottom, the layer of a light metal (\eg Na, Li, Ca, or Mg) at the top, and the thin layer of molten-salt electrolyte sandwiched in the middle. The electrolyte is selected so that it is  immiscible with the metals and conductive to positive ions of the light metal. It has the density intermediate between those of the light and heavy metals, so the entire system is stably stratified by density. The sidewalls are electrically insulated, while the horizontal top and bottom walls are conducting and serve as current collectors. The system is maintained at a temperature above the melting points of all the three materials (200 to 700$^{\circ}$C depending on the materials used). 

The energy stored in the battery is the difference between the Gibbs free energies of the light metal in its pure state and the state of alloy with the heavy metal. The battery is charged when the light metal is electrochemically reduced from the alloy and discharged when the  alloy is formed. The reactions occur in liquid state in the bottom layer in the presence of strong ($\sim$ 1 A/cm$^2$) electric currents flowing in the vertical direction.

The illustration in Fig.~\ref{fig1}a lets a fluid dynamicist to  recognize immediately that the hydrodynamic instabilities can be a major factor of the system's operation, especially when the transition from small laboratory prototypes to large commercial devices is made. The effect of the instabilities may be positive, \eg when the resulting flow in the bottom layer enhances mixing of reactants. It can also be negative, when the deformation of the interfaces between the layers becomes so strong that it causes rupture of the electrolyte layer, which means short circuit between the metals and disruption of the battery's operation. 

Several mechanisms of instability have been identified and analyzed so far including the Tayler instability 
\cite{Stefani:2011,Seilmayer:2012,Weber:2014,Herreman:2015}, the thermal convection caused by the volumetric Joule heating of the electrolyte \cite{Shen:2016,Xiang:2017} or by bottom heating \cite{Kelley:2014}, and the rolling pad instability \cite{Zikanov:2015,Weber:2016,Bojarevics:2017,Horstmann:2017}. We should also mention the electrovortex effect, which may become a significant factor under certain conditions \cite{Weber:2015}. The analysis is far from complete, but it can already be said that all these mechanisms are likely to be active in large-scale batteries. At the same time, the preliminary estimates made in \cite{Herreman:2015}  for the Tayler instability and in \cite{Shen:2016} for the Joule-heating convection show significant damping effect of the stable density stratification among the layers. It is possible, although not yet proven, that in a wide range of operational parameters, the instabilities are present but do not lead to a disruption. Rather, the instabilities lead to saturated states with weak flows and small amplitudes of interface deformation.

In this paper, we study the rolling pad instability. It is conceptually similar to the instability observed and well studied in the Hall-H\'{e}roult aluminum reduction cells (see \eg \cite{Davidson:2016}). The reduction cell is a shallow and large rectangular bath filled with molten aluminum at the bottom and molten salt electrolyte with alumina dissolved in it at the top. Nearly vertical  currents of density $\sim$ 1 A/cm$^2$ pass through the two layers and cause the desired electrochemical reduction of aluminum from its oxide. 

The reduction cell is different from the liquid metal battery in many respects: two layers instead of three, turbulent mixing of melts driven by gas bubbles and Lorentz forces, the top wall consisting of many separate carbon anodes, \etc The one important similarity is that in both the systems, the liquid domain consists of horizontal layers with vastly different electric conductivities. The conductivity of molten-salt electrolytes is about four orders of magnitude lower than that of  liquid metals. This means that even a small variation of the local thickness of the electrolyte causes large variation of local electric resistance and, thus, significantly changes the distribution of the electric currents within the system.  Lorentz forces are created by the current perturbations interacting with the magnetic field inevitably present within the cell. In unstable systems, the forces modify the flows of the melts in such a way that the interface deformation is enhanced.

In the aluminum production, the instability is viewed as a major concern. It  develops in a reduction cell when the thickness of the electrolyte is reduced below a certain cell-specific threshold and leads to sloshing waves that grow and, eventually, cause short circuit between the aluminum and the anode. The thickness has to be kept large enough, which results in energy losses to the excessive Joule heating of the electrolyte. The situation improved in the 1980s and 1990s when it was understood that the instability was caused by the interaction between the horizontal components of the current perturbations and the vertical component of the magnetic field generated by the currents flowing in the neighboring cells and electric supply lines. 

A detailed discussion of the instability mechanism can be found, \eg in  \cite{Bojarevics:1994,Sneyd:1994,Davidson:1998,Davidson:2016}. 
Briefly, as illustrated in Fig.~\ref{fig1}b, in which we should ignore the metal layer {\A } for to consider a reduction cell, the vertical current perturbations $\bm{j}_z$ form as a result of the  variation of the local thickness of the electrolyte caused by a deformation of interface. The current perturbations close within the highly conducting metal layer resulting in a horizontal component $\bm{j}_{\bot}$. The interaction between this component and the base vertical magnetic field $\bm{B}_0$ create the horizontal Lorentz force $\bm{f}$ that drives the metal in the horizontal direction perpendicular to the plane of the interface deformation. The result is a large-scale interfacial wave structure rotating around the cell - a so-called {rolling pad}.  

The mechanism by which the rotating wave grows in amplitude, and, thus, the instability develops is easiest to understand using the linearized shallow water model. In this case, the problem is reduced to a single wave equation (see \eg \cite{Davidson:1998}). Application of the Fourier expansion shows that the Lorentz forces modify the natural gravitational standing waves on the interface and introduce coupling between them. In the unstable situation, the eigenvalues of one or several couples of such waves merge to form complex-conjugate pairs, each corresponding to a rotating and exponentially growing wave. 
The shape of the horizontal cross-section of the cell determines the natural gravitational modes and, thus, strongly affects the threshold of the instability. In particular, cells of square or circular cross-section are degenerate in the sense that they  have multiple gravitational modes with equal frequencies. In the inviscid limit, such cells are unstable at arbitrarily weak electromagnetic effect. For rectangular cells, the strength of the effect must exceeds a certain not very large threshold. In the real cells, the threshold is increased by the viscosity and interfacial tension, and the situation is further complicated by the background melt flows, gas bubbles, block structure of the anode, and other factors, but the principal instability mechanism remains the same.

Fig.~\ref{fig1}b illustrates the evident fact that a similar instability can develop in a liquid metal battery. The presence of the second metal layer, in which perturbation currents $\bm{j}_{\bot}$ of opposite orientation develop (we discuss this in detail in section \ref{sec:forces}), makes the picture more complicated. The first attempt to analyze the instability was made in \cite{Zikanov:2015}. A mechanical model based on the approach developed earlier for the aluminum reduction cells in \cite{Davidson:1998} was used. In the model, the sloshing motions of the metal layers are imitated by pendulum-like oscillations of two solid metal slabs, which are independently suspended and separated  from each other by an electrolyte. The oscillations are modified and coupled to each other by the same physical mechanism as in the battery, i.e. via the Lorentz forces associated with the current perturbations caused by the local changes of the electrolyte thickness. The model is a drastic simplification of a real battery, but provides what can be viewed as a low-mode, nondissipative, linearized analogy of some aspects of the instability. 

The main result of \cite{Zikanov:2015} is the clear indication that the rolling pad instability similar to the instability in the reduction cells should be expected in the batteries. The model also suggests existence of an instability caused by the interaction between the current perturbations and the azimuthal magnetic field induced by the base current. 

The existence of the rolling pad instability in the liquid metal batteries was recently confirmed in the three-dimensional \cite{Weber:2016} and two-dimensional shallow water \cite{Bojarevics:2017} numerical simulations. Batteries of cylindrical \cite{Weber:2016} or rectangular \cite{Bojarevics:2017} shapes were considered. The typical fluid densities of a Mg-Sb battery were chosen, for which the density jump and, thus, the stabilizing effect of the buoyancy force is much larger at the lower than at the upper interface. The situation becomes particularly close to that of a reduction cell in the sense that the instability leads to noticeable deformation of only one interface. As another indication of the closeness, it has been found in  \cite{Weber:2016} that the onset of the instability is best determined by the  generalization of the stability criterion developed for the reduction cells in \cite{Sele:1977} and \cite{Urata:1985}.

This paper reports the results of a new computational study. We present a detailed derivation of the nonlinear shallow water model of a liquid metal battery. The model is then applied in a numerical study of the rolling pad instability. The goals are to \emph{(i)} further analyze the instability appearing at a single interface in the situation of strongly different density jumps and \emph{(ii)} extend the analysis to the more complex case when the density jumps are comparable to each other, and so both the interfaces are significantly deformed.   

As we discuss in detail in the model derivation in section \ref{sec:modmeth}, its validity is limited to shallow cells, in which the horizontal dimensions are much larger than the height. This limitation does not appear especially strict when we consider the main technological requirements to a large commercial battery. It must contain large quantities of both metals, have good vertical mixing in the bottom layer. The electrolyte layer must be thin and have large area in the horizontal plane, so as to maximize the reaction rate and minimize the Joule heat losses. This suggests the optimal shape of the battery in the form of a horizontally large (several m) and shallow (perhaps tens of cm) cell. The additional benefit of such a shape can be avoidance of disruptively strong Tayler and thermal convection instabilities. 


\section{Physical model and numerical method}\label{sec:modmeth}
A battery in the form of a shallow cell shown schematically in Fig.~\ref{fig1}a is considered. The cell is filled with three liquid layers: metal {\B } at the bottom, metal {\A } at the top, and electrolyte {\E } in the middle. The sidewalls are vertical and electrically insulated. The top and bottom walls are horizontal and serve as the electrodes, between which strong electric current of density $\bm{J}$ flows through the cell. 

In the following discussion, variables with the superscripts A, B, or E are associated with particular layers. Variables without superscripts either mark universal variables common for all layers (such as the horizontal length scale $L$) or, for the  sake of brevity, stand for the respective variables in each layer (\eg $H$ standing for the layer thicknesses $H^A$, $H^B$, and $H^E$).

\subsection{Assumptions made in the model}\label{sec:ass}
The model describes the behavior related to the interaction between the magnetic fields and the perturbations of electric current arising in response to the interface deformation. The physical effects leading to the other instabilities  mentioned in section \ref{sec:intro} are ignored. The justification for this approach is based on the following two arguments. The small-scale, possibly turbulent motions of melts caused by thermal convection and interfacial forces  increase the dissipative effects, but are unlikely to substantially change the physical mechanisms of the large-scale rolling pad instability. The validity of this argument is confirmed by the results obtained for the aluminum reduction cells {(see, e.g. \cite{Zikanov:2000,Moreau:1984})} and further supported in the following discussion. Furthermore, as we discuss below, the Tayler instability and the electrovortex flow are either absent or weak in the simplified shallow geometry we consider. 

The model is based on the following simplifying assumptions.
\begin{enumerate}
\item The cell is shallow in the sense that the thicknesses of all three layers are much smaller than the typical horizontal size of the cell:
\begin{equation}\label{shal}
H^{A},H^{B},H^{E}\ll L.
\end{equation}
This is the strongest of our assumptions, and  the only one that seriously limits the applicability of the model. 
We further assume that the typical horizontal wavelength $\lambda$ of the instability is large:
\begin{equation}\label{wavel}
H^{A},H^{B},H^{E}\ll \lambda\sim L.
\end{equation}
{The error of approximation of the model can be expressed in terms of the shallowness parameter}
\begin{equation}
\label{delta}
\delta\equiv \frac{H}{L} \sim \frac{H}{\lambda}\ll 1.
\end{equation}
{As we will see in the following discussion,  it is partially of the second order (e.g. for the inviscid part of the momentum transport) and partially of the first order (e.g. for the effects of viscosity or some electromagnetic effects).}
\item The metals are immiscible with the electrolyte, so sharp deformable interfaces exist between the layers.
\item Each layer is assumed to contain a  liquid of constant physical properties. This is a simplification, especially in the case of the bottom layer, where a mixture of metal {\B } with the alloy of {\A } and {\B } in liquid or, possibly, solid intermetallic state is to be found unless the battery is fully charged. 
\item The system is stably stratified by density:
\begin{equation}\label{dens}
\rho^{A}<\rho^{E}<\rho^{B}.
\end{equation}
\item The electric conductivities of the liquids satisfy
\begin{equation}\label{cond}
\sigma^{E}\ll \sigma^{A}\sim \sigma^{B}.
\end{equation}
In real batteries, $\sigma^{E}$ is about four orders of magnitude smaller than $\sigma^{A}$ and $\sigma^{B}$. 
\item The base electric current flowing through a cell with unperturbed horizontal interfaces is purely vertical and uniform, so its density is 
\begin{equation}\label{curr}
\bm{J}_0=J_0\bm{e}_z, \quad J_0=const.
\end{equation}
This means that the electrovortex effect is excluded from consideration.
\item A simplified model of the magnetic field is used. We limit the analysis to the effect of the vertical magnetic field that is generated in a battery by the currents in external circuits. The field is approximated as
\begin{equation}
\label{mf_base}
\bm{B}_0=B_0\bm{e}_z, \:\: B_0=const.
\end{equation}
The other components, in particular the magnetic field generated by the base current $\bm{J}_0$ within the cell and the perturbation field $\bm{b}$ induced by the current perturbations are not considered. 
This does not allow us to analyze full three-dimensional dynamics of the battery and excludes from consideration the Tayler instability, the effect of $\bm{b}$ on the rolling pad instability, and the possible second type instability suggested by the mechanical model \cite{Zikanov:2015}. The rationale of the approach is as follows.  Firstly, we leave the simulations of the full three-dimensional dynamics of the system to future studies. The focus of our work is on the rolling pad instability during its low-amplitude stages when it can, as an approximation, be separated from the other magnetohydrodynamic effects. A similar approach was used for analyzing the classical rolling pad instability in the reduction cells \cite{Bojarevics:1994,Sneyd:1994,Davidson:1998,Sele:1977,Urata:1985,Sun:2004,Zikanov:2000} and in the recent studies of liquid metal batteries \cite{Weber:2016,Bojarevics:2017,Horstmann:2017}. Secondly,
the Tayler instability is not anticipated in our system, since it is associated with tall cells (as shown, \eg in \cite{Rudiger:2007}, the typical axial wavelength of the instability is larger than the horizontal dimension of the cell). Thirdly, the general effect of the magnetic field perturbations $\bm{b}$ on the rolling pad instability, while present, is likely to be insignificant. This is indicated by the results obtained in \cite{Sun:2004} for the reduction cells.  Finally, a proper analysis of the instability of the second type predicted in \cite{Zikanov:2015} would require accurate evaluation of the magnetic fields induced by both the base and perturbation currents, which can only be done in the framework of the three-dimensional model that includes the interior of the cell and the adjacent conductors. This analysis is left to future studies, as well. 
\item As explained in detail in section \ref{sec:forces}, only the  electric current perturbations caused by the deformation of the interfaces are considered. 
This is justified since, as demonstrated in the three-dimensional simulations \cite{Weber:2016} and confirmed by our results presented below, in an unstable system these perturbations are typically much stronger than the electric currents induced by the melt velocities. 
 One consequence of this assumption is that the current perturbations can be represented as gradients of electric potential functions.  Full Ohm's law may need to be applied in a three-dimensional nonlinear analysis of melt flows in a battery with full three-dimensional magnetic field.
\item The melts are assumed to be at a constant temperature. The thermal convection flows are neglected despite the fact that, as discussed in \cite{Shen:2016}, they are practically unavoidable and likely to be turbulent. This is justified by the estimates (see \eg \cite{Shen:2016}) showing that the stable density stratification between the layers is sufficiently strong to prevent significant interface deformation by the convection in all, but very large batteries. Our assumption is that the convection results in small-scale, possibly turbulent flows but does not critically interfere with the large-scale rolling pad instability.
\item We now describe the base (unperturbed) state, which is designated in the following discussion by subscript 0.
The simplifying assumptions made above imply that the base state has purely vertical current (\ref{curr}) and flat interfaces (see Fig.~\ref{fig1}a):
\begin{equation}
\label{basestate}
z=\zeta^A_0=\frac{H_0^E}{2}, \: z=\zeta^B_0=-\frac{H_0^E}{2},
\end{equation}
where $H_0^E$ is the unperturbed thickness of the electrolyte layer.  The Lorentz force $\bm{J}_0 \times \bm{B}_0=0$, and the melt velocities are all zero.
This form of the base state is a simplification. In the real battery, $\bm{J}_0$ is not uniform and $\bm{B}_0$ is a three-dimensional field created by the currents within and without the cell. This and other features of the battery system, for example the thermal convection, inevitably lead to a base state with deformed interfaces and possibly turbulent melt flows. The analysis of the effect of a nontrivial base state on the instability is left to future studies. We note that the model derived below can be applied to systems with background flows and interface deformations after minor modifications (see \cite{Sun:2004,Zikanov:2000} for examples of such analysis in the case of aluminum reduction cells).
\end{enumerate}

As a final comment, we note that the model is nonlinear and can be applied to simulation of finite-amplitude as well as small-amplitude perturbations. 


\subsection{Electric current perturbations and Lorentz forces}\label{sec:forces}
In the model, we consider the Lorentz forces that arise as a result of the interface deformations, more specifically, as a result of the local change in the thickness of the electrolyte layer
\begin{equation}
\label{eettaa}
\eta(x,y,t)=H^E(x,y,t)-H^E_0=\zeta^A(x,y,t)-\zeta^B(x,y,t)-H^E_0
\end{equation}
(see Fig.~\ref{fig1}a). The following derivation is a generalization of the derivation used for the aluminum reduction cells (see \eg \cite{Sun:2004}). The starting point is the approximation based on the relation (\ref{cond}). 
 {Due to the much higher electric conductivities of the metals,  the distribution of currents in the electrolyte can be accurately approximated as corresponding to equipotential interfaces $z=\zeta^A(x,y,t)$ and $z=\zeta^B(x,y,t)$.  Combined with the shallowness of the electrolyte layer, this allows us to approximate the perturbed current in this layer as vertical (following the shortest path through a resistive medium) and, due to their zero divergence, $z$-independent:}
\begin{equation}
\label{electr_pert}
\left(J_0+j_z^E(x,y,t)\right)\bm{e}_z
\end{equation}
and to express it via the Ohm's law as
\begin{equation}
\label{jze_ohm}
J_0+j_z^E=\sigma^E \frac{\Phi^A-\Phi^B}{H^E} \textrm{ (perturbed state)}, \: J_0=\sigma^E \frac{\Phi^A_0-\Phi^B_0}{H^E_0} \textrm{ (base state)},
\end{equation}
where $\Phi^A$ and $\Phi^B$ are the values of the electric potential at the top and bottom interfaces.  

We now set $\Phi^B=\Phi^B_0=0$ and notice that the total electric resistance of the electrolyte changes by the quantity $\sim \left(\eta/H^E_0\right)^2$. Either the net current through the cell or the voltage drop has to change  when $\eta(x,y,t)$ is not infinitesimal. Following the approach of \cite{Zikanov:2000}, we assume that  the voltage drop adjusts as $\Phi^A=C(t)\Phi^E_0$ while the total current remains the same (this corresponds, e.g., to the situation when the considered cell is one of many connected in parallel):
\begin{equation}
\label{totcur}
\langle J_0+j_z^E\rangle \equiv \int_0^{L_y}\int_0^{L_x}\left(J_0+j_z^E \right) dx dy=\langle J_0\rangle.
\end{equation}
This allows us to determine the current perturbations as
\begin{equation}
\label{jez}
j_z^E=\sigma^E\left(\frac{C(t)\Phi^A_0}{H^E}-\frac{\Phi^A_0}{H_0^E} \right)=J_0\left(\frac{C(t)H_0^E}{H^E} -1\right),
\end{equation}
where the non-dimensional adjustment coefficient is 
\begin{equation}
\label{ccoeff}
C(t)=\frac{\langle J_0\rangle}{\langle J_0H_0^E/H^E \rangle}.
\end{equation}
{At infinitesimal perturbations $\eta\ll H_0^E$, the coefficient becomes $C\approx 1$, and the expression (\ref{jez}) can be approximated using the first-order Taylor expansion in $\eta/H_0^E$ as}
\begin{equation}
\label{jze_inf}
j_z^E \approx -J_0\frac{\eta}{H_0^E}.
\end{equation}

As the next step, we assume that the perturbation currents completely close within the metal layers {\A } and \B. For the currents in the layer \A, we integrate 
\begin{equation}
\label{divnol}
\nabla \cdot \bm{j^A}=0
\end{equation}
from $z=\zeta^A$ to the top boundary $z=z_t$. {Using the Leibnitz integration rule, the condition that the vertical perturbation current at $z=z_t$ is zero, and the condition of continuity of normal current at the interface}
\begin{equation}
\label{normcont}
\left. j_z^E\right|_{z=\zeta^A}=\left. j_z^A\right|_{z=\zeta^A}-\left. j_x^A\right|_{z=\zeta^A}\frac{\partial \zeta^A}{\partial x}-\left. j_y^A\right|_{z=\zeta^A}\frac{\partial \zeta^A}{\partial y}
\end{equation}
{we obtain}
\begin{equation}
\label{dive}
\frac{\partial J_x^A}{\partial x}+\frac{\partial J_y^A}{\partial y}=j_z^E,
\end{equation}
where
\begin{equation}
\label{jjxy}
J_x^A=\int_{\zeta^A}^{z_t} j_x^Adz, \: J_y^A=\int_{\zeta^A}^{z_t} j_y^Adz
\end{equation}
are the vertically integrated horizontal  components of the perturbation current  in the layer \A.

{The solution for $\bm{J}^A$ can be found in a straightforward manner if we assume that }
\begin{equation}
\label{curlJ}
\nabla_{\bot} \times \bm{J}=0 
\end{equation}
{and express (\ref{dive}) as}
\begin{equation}
\label{potential}
\bm{J}^A_{\bot}=\nabla_{\bot} \Psi^A, \:\: \nabla_{\bot}^2 \Psi^A = j_z^E,
\end{equation}
where $\bot$ marks the horizontal $(x,y)$ plane. 

{To verify (\ref{curlJ}), we consider that the three-dimensional current perturbations are curl-free (see assumption 8 in section \ref{sec:ass}). Integration of $\nabla\times \bm{j}^A=0$ from $z=\zeta^A$ to $z=z_t$ gives}
\begin{equation}
\label{curlJ2}
\nabla_{\bot}\times \bm{J}^A_{\bot}=\left. j_x^A\right|_{z=\zeta^A}\frac{\partial \zeta^A}{\partial y}-\left. j_y^A\right|_{z=\zeta^A}\frac{\partial \zeta^A}{\partial x}.
\end{equation}
{The right-hand side of (\ref{curlJ2}) is of the second order in terms of the perturbation amplitude, so it can be neglected when linear stability to infinitesimal perturbations is analyzed. The situation is more complex for perturbations of finite amplitude. The individual terms in the right-hand and left-hand sides of (\ref{curlJ2}) are of the same order in terms of the shallowness parameter (\ref{delta}). This follows from the estimates $j^A\sim HJ^A$, $\zeta^A\sim H$, and $\partial/\partial x\sim\partial/\partial y\sim \lambda^{-1}$. A closer consideration, however, reveals that the entire right-hand side of (\ref{curlJ2}) is likely to be negligibly small. It is a cross-product of two two-dimensional vector fields: $\left.\bm{j}^A\right|_{z=\zeta^A}$ and $\nabla_{\bot}\zeta^A$. The currents $\left.\bm{j}^A\right|_{z=\zeta^A}$ are flowing in the direction parallel to the local steepest gradient of the perturbation current $j_z^E$ injected into the metal. This direction is, in turn, aligned with the direction of $\nabla_{\bot}\eta$. This follows from (\ref{jez}) and is immediately seen in the linearized version (\ref{jze_inf}). If the density differences between the metals and the electrolyte satisfy $\rho^E-\rho^A \ll \rho^B-\rho^E$, so only the upper interface is significantly deformed (see section \ref{sec:padsingle}), $\nabla_{\bot}\eta\approx \nabla_{\bot}\zeta^A$. The situation is less certain when the density jumps and, thus, the deformations of the two interfaces are comparable (see section \ref{sec:paddouble}). For this case we can argue that the interfacial waves are nearly anti-symmetrically coupled (see \cite{Horstmann:2017} and our results in section \ref{sec:paddouble}), so $\nabla_{\bot}\eta$ and $\nabla_{\bot}\zeta^A$ are nearly parallel.} 

{In summary, we see that the vectors $\left.\bm{j}^A\right|_{z=\zeta^A}$ and $\nabla_{\bot}\zeta^A$ are perfectly or nearly parallel in the flow. We can neglect the right-hand side of (\ref{curlJ2}) and consider the field of integrated currents $\bm{J}^A$ as curl-free. While highly plausible and certainly correct in many cases, this approximation may become inaccurate in some situations, in which case the model becomes only valid for infinitesimal perturbations.}

For the vertically integrated  perturbation currents in the layer \B, we can derive similar relations with $-j_z^E$ in the right-hand side of the Poisson equation or simply observe that the  conservation of electric charge requires 
\begin{equation}
\label{chargecons}
\bm{J}^B_{\bot}=-\bm{J}^A_{\bot}.
\end{equation}

In our case of purely vertical magnetic field, the Lorentz forces are zero in the layer {\E } and purely horizontal in the layers {\A } and \B. The vertically integrated forces are
\begin{equation}
\label{ab-forces}
\bm{F}_{\bot}^A=\bm{J}^A_{\bot}\times\left(B_0\bm{e}_z\right), \:\: \bm{F}_{\bot}^A=-\bm{F}_{\bot}^B.
\end{equation}

In the general case of a three-dimensional magnetic field $\bm{B}$ we would need to use the general three-dimensional expression $\bm{F}=\bm{J}\times \bm{B}$. In addition to the horizontal currents (\ref{potential}), (\ref{chargecons}), the field of the perturbation currents $\bm{J}$ would include the vertically integrated vertical components 
\begin{equation}
\label{jvert}
J_z^E=H^E j_z^E, \:\: J_z^A=\frac{H^A}{2}j_z^E, \:\:  J_z^B=\frac{H^B}{2}j_z^E,
\end{equation}
where the last two expressions are approximations with the error $\sim\delta^2$.


\subsection{Shallow water approximation}\label{sec:model}
{The derivation of the two-dimensional model follows the principal steps used for shallow flows in hydrology. There are differences, in particular the presence of Lorentz forces and immiscible layers of distinct physical properties in our case. Similarly to the approach applied in  \cite{Sun:2004} to the aluminum reduction cells, the shallow water equations in the conservative de St. Venant form are derived and used. The main advantage of this approximation over the simpler non-conservative shallow water model is that its error in representing the inviscid flow's dynamics is of the second rather than first order in terms of the shallowness parameter (\ref{delta}).}
As discussed in \cite{Sun:2004}, the  two models produce nearly identical solutions in the case of weak interface deformations, but diverge noticeably for the nonlinear states, where the deformations are strong.

The derivation starts with the full three-dimensional momentum equations for flows in each layer, with the liquids treated as incompressible, inviscid and having constant and uniform physical properties. The effect of viscosity will be introduced at a later stage. For the sake of generality and possible future use, the equations are derived for the case of a general three-dimensional Lorentz force that would exist in a cell with a general three-dimensional magnetic field.
With the purely vertical currents in the electrolyte and three-dimensional currents in the metals, the general force field has the form
\begin{eqnarray}
\bm{f}^A & = & \bm{f}_{\bot}^A+f_z^A\bm{e}_z,\\
\bm{f}^E & = & \bm{f}_{\bot}^E,\\
\bm{f}^B & = & \bm{f}_{\bot}^B+f_z^B\bm{e}_z.
\end{eqnarray}
The reduction to the actually considered in this paper case of a purely vertical magnetic field, in which the force is two-dimensional, will be done later.

The boundary conditions are those of zero velocity at all walls. At the interfaces $z=\zeta^A(x,y,t)$ and $z=\zeta^B(x,y,t)$, continuity of pressure and the kinematic condition 
\begin{equation}\label{interface}
\frac{\partial \zeta}{\partial t} +u\frac{\partial \zeta}{\partial x}+v\frac{\partial \zeta}{\partial y}  = w,
\end{equation}
where $u$, $v$, and $w$ are the velocity components, are required.  

The incompressibility implies the estimate of the vertical velocity
\begin{equation}
\label{vervel}
w\sim\delta u\sim\delta v.
\end{equation}
Applying it to the $z$-momentum equation and dropping all the terms $\sim \delta^2$ we obtain, in each layer:
\begin{equation}
\label{hydro}
\frac{\partial p}{\partial z}=-\rho g+f_z.
\end{equation}
Next, we introduce the pressure distribution at the mid-plane of the electrolyte layer $z=0$
\begin{equation}
\label{pre0}
p_0(x,y,t)\equiv p(x,y,z=0,t)
\end{equation}
and integrate (\ref{hydro}) in $z$. The resulting pressure distributions satisfying the requirement of continuity at the interfaces are:
\begin{eqnarray}
\label{pre3d-1} p^A(x,y,z,t) & = & p_0+(\rho^A-\rho^E)g\zeta^A-\rho^Agz+\int_{\zeta^A}^z f_z^Ad\xi,\\
\label{pre3d-2}  p^E(x,y,z,t) & = & p_0-\rho^Egz,\\
\label{pre3d-3} p^B(x,y,z,t) & = & p_0+(\rho^B-\rho^E)g\zeta^B-\rho^Bgz+\int_{\zeta^B}^z f_z^Bd\xi.
\end{eqnarray} 
They are substituted into the horizontal momentum equations written in conservation form for each layer. Together with the incompressibility conditions, the equations are:
\begin{eqnarray}
\frac{\partial u}{\partial x}+\frac{\partial v}{\partial y}+\frac{\partial w}{\partial z} & = & 0,\\
\frac{\partial u}{\partial t}+\frac{\partial}{\partial x}\left(u^2\right)+\frac{\partial}{\partial y}\left(uv\right)+\frac{\partial}{\partial z}\left(uw\right) & = & -\frac{1}{\rho}\frac{\partial p}{\partial x} +\frac{1}{\rho}f_x,\\
\frac{\partial v}{\partial t}+\frac{\partial}{\partial x}\left(uv\right)+\frac{\partial}{\partial y}\left(v^2\right)+\frac{\partial}{\partial z}\left(vw\right) & = & -\frac{1}{\rho}\frac{\partial p}{\partial y} +\frac{1}{\rho}f_y.
\end{eqnarray} 
{The approaches of the simple non-conservative shallow water model and the conservative de St.~Venant model diverge at this point.} In the former, we would just assume that $w=0$ and all the other variables are $z$-independent. This would be equivalent to neglecting all but the zero-order terms in the $\delta$-expansions of the flow fields. Instead, we follow the  routine of the second order in the shallowness parameter (\ref{delta}) based on the integration of the equations vertically across each layer. 

The integration requires the use of the Leibniz rule for the $x$-, $y$-, and $t$-derivatives, the zero-velocity conditions at $z=z_b$ and $z=z_t$, and the interface conditions (\ref{interface}). 
We use the velocity fluxes
\begin{equation}
\label{fluxes}
\bm{U}^A(x,y,t)=\int_{\zeta^A}^{z_t}\bm{u}^Adz, \: \bm{U}^B(x,y,t)=\int_{z_b}^{\zeta^B}\bm{u}^Bdz, \: \bm{U}^E(x,y,t)=\int_{\zeta^B}^{\zeta^A}\bm{u}^Edz,
\end{equation}
and the integrated horizontal components of the Lorentz forces
\begin{equation}
\label{intforces}
\bm{F}^A_{\bot}(x,y,t)=\int_{\zeta^A}^{z_t}\bm{f}^A_{\bot}dz, \: \bm{F}_{\bot}^B(x,y,t)=\int_{z_b}^{\zeta^B}\bm{f}^B_{\bot}dz, \: \bm{F}_{\bot}^E(x,y,t)=\int_{\zeta^B}^{\zeta^A}\bm{f}^E_{\bot}dz.
\end{equation}
We also need the vertically integrated terms of the horizontal gradients of the pressure fields (\ref{pre3d-1})-(\ref{pre3d-3}), of which only the Lorentz force terms are non-trivial.
For them, we apply the Leibniz rule and the approximation to the second order in $\delta$ to obtain
\begin{eqnarray}
\label{prevert1}
\int_{\zeta^A}^{z_t}\nabla_{\bot}\left(\int_{\zeta^A}^z f_z^A d\xi \right)dz & \approx & \nabla_{\bot}\left(\frac{H^AF_z^A}{2}\right), \\ 
\label{prevert1} \int_{z_b}^{\zeta^B}\nabla_{\bot}\left(\int_{\zeta^B}^z f_z^B d\xi \right)dz & \approx & - \nabla_{\bot}\left(\frac{H^BF_z^B}{2}\right),
\end{eqnarray}
 where 
\begin{equation}
\label{intzforces}
F_z^A=\int_{\zeta^A}^{z_t} f_z^A dz, \: F_z^B=\int_{z_b}^{\zeta^B} f_z^B dz.
\end{equation}
In the integration of the nonlinear terms of the momentum equations, we utilize the second order approximations, such as
\begin{equation}
\label{nonliapp}
\int_{\zeta^A}^{z_t}\left(u^A\right)^2dz \approx \frac{1}{H^A}\left(U^A\right)^2. 
\end{equation}

The final momentum equations are
\begin{eqnarray}
\nonumber  \frac{D \bm{U}^A}{D t} & = & -\frac{H^A}{\rho^A}{\nb}  {p_0} -\left(1-\frac{\rho^E}{\rho^A}\right)H^Ag\nb\zeta^A-\frac{1}{2\rho^A}\nb\left(F_z^AH^A\right)\\
\label{mom1} & & +\frac{\bm{F}_{\bot}^A}{\rho^A}+\bm{\tau}^A,\\
\label{mom2} \frac{D \bm{U}^E}{D t} & = & -\frac{H^E}{\rho^E}\nb  p_0 +\frac{\bm{F}_{\bot}^E}{\rho^E}+\bm{\tau}^E,\\
\nonumber  \frac{D \bm{U}^B}{D t} & = & -\frac{H^B}{\rho^B}\nb  p_0 -\left(1-\frac{\rho^E}{\rho^B}\right)H^Bg\nb\zeta^B+\frac{1}{2\rho^B}\nb\left(F_z^BH^B\right)\\
\label{mom3} & & +\frac{\bm{F}_{\bot}^B}{\rho^B}+\bm{\tau}^B,
\end{eqnarray} 
where $\bm{\tau}^A$, $\bm{\tau}^B$, and $\bm{\tau}^E$ are the viscous friction terms, which we introduce here for the first time and discuss in detail below, and  the material derivative in the left-hand side of each equation has the $x$- and $y$-components
\begin{eqnarray}
\label{mater1} \frac{D U}{D t} & = & \frac{\partial U}{\partial t}+\frac{\partial}{\partial x}\left(\frac{U^2}{H}\right)+\frac{\partial}{\partial y}\left(\frac{UV}{H}\right),\\
\label{mater2}  \frac{D V}{D t} & = & \frac{\partial V}{\partial t}+\frac{\partial}{\partial x}\left(\frac{UV}{H}\right)+\frac{\partial}{\partial y}\left(\frac{V^2}{H}\right).
\end{eqnarray}

The incompressibility condition is also integrated using the Leibniz rule and the interface and wall boundary conditions. This leads to
\begin{eqnarray}
\label{int-1} \frac{\partial \zeta^A}{\partial t} & = & \nabla_{\bot} \cdot \bm{U}^A,\\
\label{int-2} \frac{\partial \zeta^B}{\partial t} & = & -\nabla_{\bot} \cdot \bm{U}^B,\\
\label{int-3} \frac{\partial \zeta^A}{\partial t}-\frac{\partial \zeta^B}{\partial t} & = & -\nabla_{\bot} \cdot \bm{U}^E.
\end{eqnarray} 
In our model, (\ref{int-1}) and (\ref{int-2}) are applied to determine the evolution of interfaces. The equation (\ref{int-3}) is replaced by the combination of  (\ref{int-1})-(\ref{int-3}) expressing the global conservation of mass
\begin{equation}
\label{masscons}
\nabla_{\bot} \cdot \left(\bm{U}^A+\bm{U}^B+\bm{U}^E \right)=0.
\end{equation}

The approximation of the viscous friction terms follows the approach commonly applied in hydrology. We retain the horizontal components of the Laplacian and approximate the friction at the horizontal walls and interfaces by finite differences. The  friction terms in (\ref{mom1})--(\ref{mom3}) are: 
\begin{eqnarray}
\label{visc1} \bm{\tau}^A & = & \frac{\mu^A}{\rho^A}\nabla_{\bot}^2\bm{U}^A+\frac{1}{\rho^A}\left[-\frac{\mu^E+\mu^A}{H^E+H^A}\left(\frac{\bm{U}^A}{H^A}-\frac{\bm{U}^E}{H^E}\right) - \frac{2\mu^A\bm{U}^A}{\left(H^A\right)^2}\right] \\
\label{visc2} \bm{\tau}^B & = & \frac{\mu^B}{\rho^B}\nabla_{\bot}^2\bm{U}^B+\frac{1}{\rho^B}\left[-\frac{\mu^B+\mu^E}{H^B+H^E}\left(\frac{\bm{U}^B}{H^B}-\frac{\bm{U}^E}{H^E}\right) - \frac{2\mu^B\bm{U}^B}{\left(H^B\right)^2}\right] \\
\nonumber \bm{\tau}^E & = & \frac{\mu^E}{\rho^E}\nabla_{\bot}^2\bm{U}^E+\frac{1}{\rho^E}\left[\frac{\mu^E+\mu^A}{H^E+H^A}\left(\frac{\bm{U}^A}{H^A}-\frac{\bm{U}^E}{H^E}\right) \right.\\
\label{visc3} & &  +\left.\frac{\mu^B+\mu^E}{H^B+H^E}\left(\frac{\bm{U}^B}{H^B}-\frac{\bm{U}^E}{H^E}\right)\right]. 
\end{eqnarray}
This friction model is decisively simple. In particular, the coefficients $\mu$ are the molecular dynamic viscosity coefficients of the respective liquids. No attempt is made to account for the possible small-scale turbulence  by using eddy viscosities or applying the models derived for turbulent open channel flows. The main justification of this approach is that, as expected for large-wavelength instabilities and confirmed by our results  
in section \ref{sec:results}, the effect of the viscous friction terms on the evolution of the large-scale modes of the rolling pad instability is limited to minor quantitative changes.

In the computational analysis presented in section \ref{sec:results}, the equations are simplified for the case of a purely vertical magnetic field by setting all the vertical force components $F_z$ and the forces in the electrolyte $\bm{F}_{\bot}^E$ to zero and using the expressions (\ref{ab-forces}) for the remaining horizontal components. The perturbation electric currents are determined as described in section \ref{sec:forces}.

\subsection{Equations in non-dimensional form}\label{sec:nondim}
As a physical system, even the simplified battery considered in this paper is described by a large number of parameters: horizontal dimensions, thicknesses and liquid properties of the three layers, base electric current and magnetic field. Converting the problem into non-dimensional form reduces this number, but still leaves many non-dimensional groups. Furthermore, since the battery is a real device, each physical parameter should be allowed to vary only within some practically meaningful limits. In such a situation, it is convenient to perform the analysis using dimensional variables. At the same time, as we will see in section \ref{sec:results}, the non-dimensional form of the problem helps to understand the physics of the instability. Our discussion takes the hybrid approach. The simulation cases are identified using  dimensional parameters, while the discussion is conducted primarily in terms of non-dimensional quantities. The non-dimensionalization is presented in this section.

We take the larger horizontal dimension of the cell, say, $L=L_x$ in a rectangular cell, as the length scale, the unperturbed current $J_0$ as the scale for the density of electric current, the constant vertical component $B_0$ as the scale of the magnetic field, and the physical properties of one liquid (we take the electrolyte) $\rho^E$, $\nu^E=\mu_E/\rho_E$ as the scales of density and kinematic viscosity. As the velocity scale, we use the typical velocity of a large-scale gravitational wave on the upper interface:
\begin{equation}
\label{grav_wave}
U_0=\left[ \frac{\left(\rho^E-\rho^A\right)g}{\rho^A\left(H_0^A\right)^{-1} + \rho^E\left(H_0^E\right)^{-1}} \right]^{1/2}.
\end{equation}
This choice, also practiced for the aluminum reduction cells (see, \eg \cite{Davidson:1998}), is consistent with the physical nature of the instability.
The typical scales of time and pressure are  $L/U_0$ and $\rho^EU_0^2$, respectively. The typical scales for the Lorentz force and the electric potential $\Psi^A$ are $J_0B_0L$ and $J_0L^2$, respectively.

Substituting the scaled variables into the equations of sections \ref{sec:model} and \ref{sec:forces} and performing the  non-dimensionalization procedure, we obtain the final non-dimensional system. It includes the momentum equations written for each of the three layers, the equations of the interface deformations, and the incompressibility condition:
\begin{eqnarray}
\nonumber & &\frac{D \bm{U}^A}{D t}  =  -\frac{H^A}{\gamma_{\rho}^A}\nb  p_0 -\left(1-\frac{1}{\gamma_{\rho}^A}\right)\frac{H^A}{\Fr^2}\nb\zeta^A-
\frac{\epsilon}{\gamma_{\rho}^A}\nb\left(\frac{F_z^AH^A}{2}\right)\\
\label{mom1nd}  & & +\frac{\epsilon}{\gamma_{\rho}^A}\bm{F}_{\bot}^A+\bm{\tau}^A,  \\
\label{mom2nd} & &\frac{D \bm{U}^E}{D t}  = -H^E\nb  p_0 +\epsilon \bm{F}_{\bot}^E+\bm{\tau}^E,  \\
\nonumber & &\frac{D \bm{U}^B}{D t}  =  -\frac{H^B}{\gamma_{\rho}^B}\nb  p_0 -\left(1-\frac{1}{\gamma_{\rho}^B}\right)\frac{H^B}{\Fr^2} \nb\zeta^B+
\frac{\epsilon}{\gamma_{\rho}^B}\nb\left(\frac{F_z^BH^B}{2}\right)\\
\label{mom3nd} & & +\frac{\epsilon}{\gamma_{\rho}^A} \bm{F}_{\bot}+\bm{\tau}^B, \\
\label{int-1nd} & & \frac{\partial \zeta^A}{\partial t}  =  \nabla_{\bot}\cdot \bm{U}^A,\\
\label{int-2nd} & & \frac{\partial \zeta^B}{\partial t}  =  -\nabla_{\bot}\cdot \bm{U}^B,\\
\label{massconsnd} & & \nabla_{\bot}\cdot\left(\bm{U}^A+\bm{U}^B+\bm{U}^E \right)=0,
\end{eqnarray} 
where all the variables are now non-dimensional and the  material derivatives in the left-hand sides of (\ref{mom1nd})-(\ref{mom3nd}) are expressed as in (\ref{mater1})-(\ref{mater2}). The non-dimensional friction terms are
\begin{eqnarray}
\label{visc1nd} \bm{\tau}^A & = & \frac{\gamma_{\nu}^A}{\Rey}\nabla_{\bot}^2\bm{U}^A+\frac{1}{\Rey}\left[-\frac{\left({\gamma_{\rho}^A}\right)^{-1}+\gamma_{\nu}^A}{H^E+H^A}\left(\frac{\bm{U}^A}{H^A}-\frac{\bm{U}^E}{H^E}\right) - \frac{2\gamma_{\nu}^A\bm{U}^A}{\left(H^A\right)^2}\right],\\
\label{visc2nd} \bm{\tau}^B & = & \frac{\gamma_{\nu}^B}{\Rey}\nabla_{\bot}^2\bm{U}^B+\frac{1}{\Rey}\left[-\frac{\left({\gamma_{\rho}^B}\right)^{-1}+\gamma_{\nu}^B}{H^B+H^E}\left(\frac{\bm{U}^B}{H^B}-\frac{\bm{U}^E}{H^E}\right) - \frac{2\gamma_{\nu}^B\bm{U}^B}{\left(H^B\right)^2}\right] \\
\label{visc3nd} \bm{\tau}^E & = & \frac{1}{\Rey}\nabla_{\bot}^2\bm{U}^E+\frac{1}{\Rey}\left[\frac{1+\gamma_{\nu}^A\gamma_{\rho}^A}{H^E+H^A}\left(\frac{\bm{U}^A}{H^A}-\frac{\bm{U}^E}{H^E}\right) + \frac{1+\gamma_{\rho}^B\gamma_{\nu}^B}{H^B+H^E}\left(\frac{\bm{U}^B}{H^B}-\frac{\bm{U}^E}{H^E}\right)\right]. 
\end{eqnarray}

The  Lorentz forces are expressed in terms of the non-dimensional variables as 
\begin{eqnarray}
\label{crpr} j_z^E & = & C(t)\frac{H_0^E}{H^E}-1, \: C(t)=\frac{A}{\langle \tilde{H}_0^E/H^E\rangle}\\
\bm{J}^E & = & J_z^E\bm{e}_z, \: J_z^E=H^E j_z^E,\\
J_z^A & = & \frac{H^A}{2}j_z^E, \: J_z^B = \frac{H^B}{2}j_z^E,\\
\label{poisnd} \nb^2\Psi^A & = & j_z^E, \\
\label{curnd} \bm{J}^A_{\bot} & = & \nb \Psi^A, \: \bm{J}^B_{\bot}=-\bm{J}^A_{\bot},\\
\bm{F} & = & \bm{J}\times \bm{B}_0,
\end{eqnarray}
where $A$ is the non-dimensional area of the horizontal cross-section of the cell. 
For the system analyzed in this paper, the non-dimensional magnetic field is $\bm{B}_0=\bm{e}_z$, so only the horizontal components of the Lorentz force are non-zero, and the model for the current perturbations reduces to (\ref{crpr}), (\ref{poisnd}), (\ref{curnd}).

The boundary conditions include the no-slip and perfect electric insulation conditions at the walls:
\begin{equation}
\label{bcbc}
\bm{U}^A=\bm{U}^B=\bm{U}^E=0, \: \frac{\partial \Psi^A}{\partial n}=0 \textrm{ at sidewalls}.
\end{equation}

The non-dimensional parameters of the problem are the Froude number
\begin{equation}
\label{froude}
\Fr\equiv \frac{U_0}{\left(Lg\right)^{1/2}},
\end{equation}
the electromagnetic parameter 
\begin{equation}
\label{elmpar}
\epsilon \equiv \frac{J_0B_0L}{\rho^EU_0^2},
\end{equation}
the Reynolds number
\begin{equation}
\label{reynolds}
\Rey\equiv \frac{U_0L}{\nu^E},
\end{equation}
and the ratios of densities and kinematic viscosities
\begin{equation}
\label{densrat}
\gamma_{\rho}^A\equiv \frac{\rho^A}{\rho^E}, \: \gamma_{\rho}^B\equiv \frac{\rho^B}{\rho^E}, \gamma_{\nu}^A\equiv \frac{\nu^A}{\nu^E}, \: \gamma_{\nu}^B\equiv \frac{\nu^B}{\nu^E}.
\end{equation}
We should also add the  parameters defining the cell's geometry: the non-dimensional unperturbed layer thicknesses $\tilde{H}_0^A$, $\tilde{H}_0^B$, $\tilde{H}_0^E$, and the parameter describing the shape in the horizontal plane, for example, the aspect ratio 
\begin{equation}
\label{pargam}
\Gamma=L_x/L_y
\end{equation}
 for a rectangular cell.

\subsection{Numerical method}\label{sec:num}
The problem is solved numerically using a finite-difference scheme similar to the scheme applied in \cite{Zikanov:2000} and \cite{Sun:2004} to the shallow water equations for aluminum reduction cells. The time discretization is based on the explicit time-splitting (projection) scheme of the third order. Each time step includes 
\begin{enumerate}
\item Solution of the Poisson equation (\ref{poisnd}) and computation of the electric current perturbations and Lorentz forces, 
\item Advancement of the time-evolution equations (\ref{mom1nd})-(\ref{int-2nd}) according to the stiffly-stable scheme of \cite{Karniadakis:1991}:
\begin{equation}
\label{scheme3}
f_*=\frac{6}{11}\left[3f^n-\frac{3}{2}f^{n-1}+\frac{1}{3}f^{n-2}+\Delta t\left( 3q^n-3q^{n-1}+q^{n-2} \right)\right],
\end{equation}
where $f$ stands for a velocity flux or interface deformation and $q$ for the respective right-hand side of the evolution equation. For the momentum equations, the pressure gradient term is not included into the right-hand side, and $f_*$ is an intermediate velocity flux. For the interface equations, $f_*$ is the value at the next time step.
\item Solution of the pressure equation
\begin{equation}
\label{prespois}
\nabla_{\bot}\cdot \left[\left(\frac{H^A}{\gamma_{\rho}^A}+\frac{H^B}{\gamma_{\rho}^B}+H^E \right)\nabla_{\bot}p_0 \right]=\frac{1}{\dt}\nabla_{\bot}\cdot\left(\bm{U}_*^A+\bm{U}_*^B+\bm{U}_*^E \right)
\end{equation}
with zero normal gradients of $p_0$ at the sidewalls as the boundary condition.
\item Correction of the velocity fluxes, in each layer, according to
\begin{equation}
\label{corrpre}
\bm{U}^{n+1}=\bm{U}_*-\frac{6\dt}{11}\frac{H}{\gamma_{\rho}}\nabla_{\bot}p_0
\end{equation}
and enforcement of the no-slip boundary conditions.
\end{enumerate}

The spatial discretization is of the second order of approximation and uses central differences implemented on a staggered grid. The Poisson equation (\ref{poisnd}) and the non-separable elliptic equation (\ref{prespois}) are solved using the multi-grid algorithms of the library Mudpack \cite{Mudpack}.

A grid sensitivity study was carried out. It has been found that the large-scale character of the instability allows us to obtain accurate results on moderately fine grids. For example, for a battery of rectangular cross-section with the aspect ratio $\Gamma=2.0$, the grid of $128\times 64$ points is certainly sufficient, while the grid of $64\times 32$ points provides the results, which are  generally correct with only minor quantitative inaccuracy. 

The analysis consists of multiple simulations, each reproducing the flow's evolution for a certain set of parameters. Each simulation begins with zero melt velocities and the flat interfaces (\ref{basestate}) perturbed randomly with the amplitude of $10^{-5}H^E_0$. The simulation continues until the maximum deformation of one of the interface exceeds $0.5H^E_0$ or the growing perturbations reach finite-amplitude saturation, or, in the stable case, for not less than 3000 non-dimensional time units {$L_x/U_0$, with $U_0$ given by (\ref{grav_wave}). }


\section{Results}\label{sec:results}

The analysis consists of two parts different from each other by the relation between the density jumps across the interfaces $\Delta \rho^A=\rho^E-\rho^A$ and $\Delta \rho^B=\rho^B-\rho^E$.  The simpler case of $\Delta \rho^A\ll \Delta \rho^B$ or $\Delta \rho^B\ll \Delta \rho^A$, in which the significant deformation only occurs at the interface with the smaller jump is considered in section \ref{sec:padsingle}. The main goal is to continue the analysis of \cite{Weber:2016,Bojarevics:2017} toward a more accurate evaluation of the extent, to which the predictions developed for the aluminum reduction cells apply to the batteries. The more complex case when $\Delta \rho^A\sim \Delta \rho^B$, so both the interfaces are significantly deformed, is discussed in section \ref{sec:paddouble}.


\subsection{Single interface deformation}\label{sec:padsingle}

\begin{figure}
\begin{center}
\includegraphics[width=1.0\textwidth]{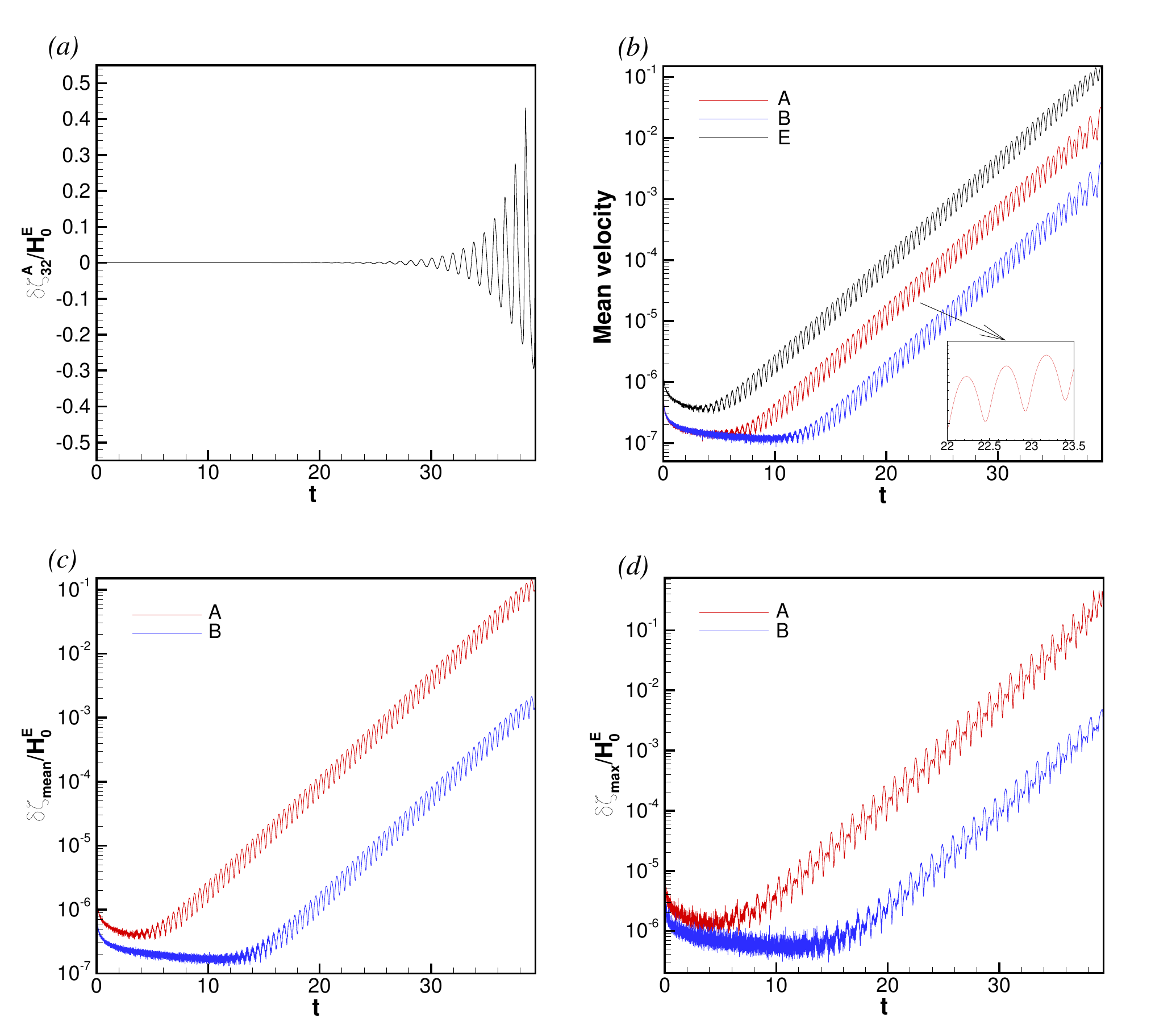}
\caption{Example of the rolling pad instability in the case when only the upper interface is significantly deformed. (see text for the system's parameters). Deformation of the upper interface at the grid point $(x_3,y_2)$ \emph{(a)}, space-averaged rms of fluid velocities $U/H$ in each layer \emph{(b)}, space-averaged rms of interface deformations \emph{(c)}, and maximum interface deformations \emph{(d)} are shown as functions of time for the entire simulated flow evolution. The interface deformations are scaled by the unperturbed electrolyte thickness $H_0^E$. Note that two states with interface deformations $\pm\delta \zeta$ have the same values of the space-averaged quantities, so the curves in  \emph{(b)} and \emph{(c)} oscillate with the period $T/2$.}
\label{fig:singleoutput}
\end{center}
\end{figure}

\begin{figure}
\begin{center}
\includegraphics[width=1.0\textwidth]{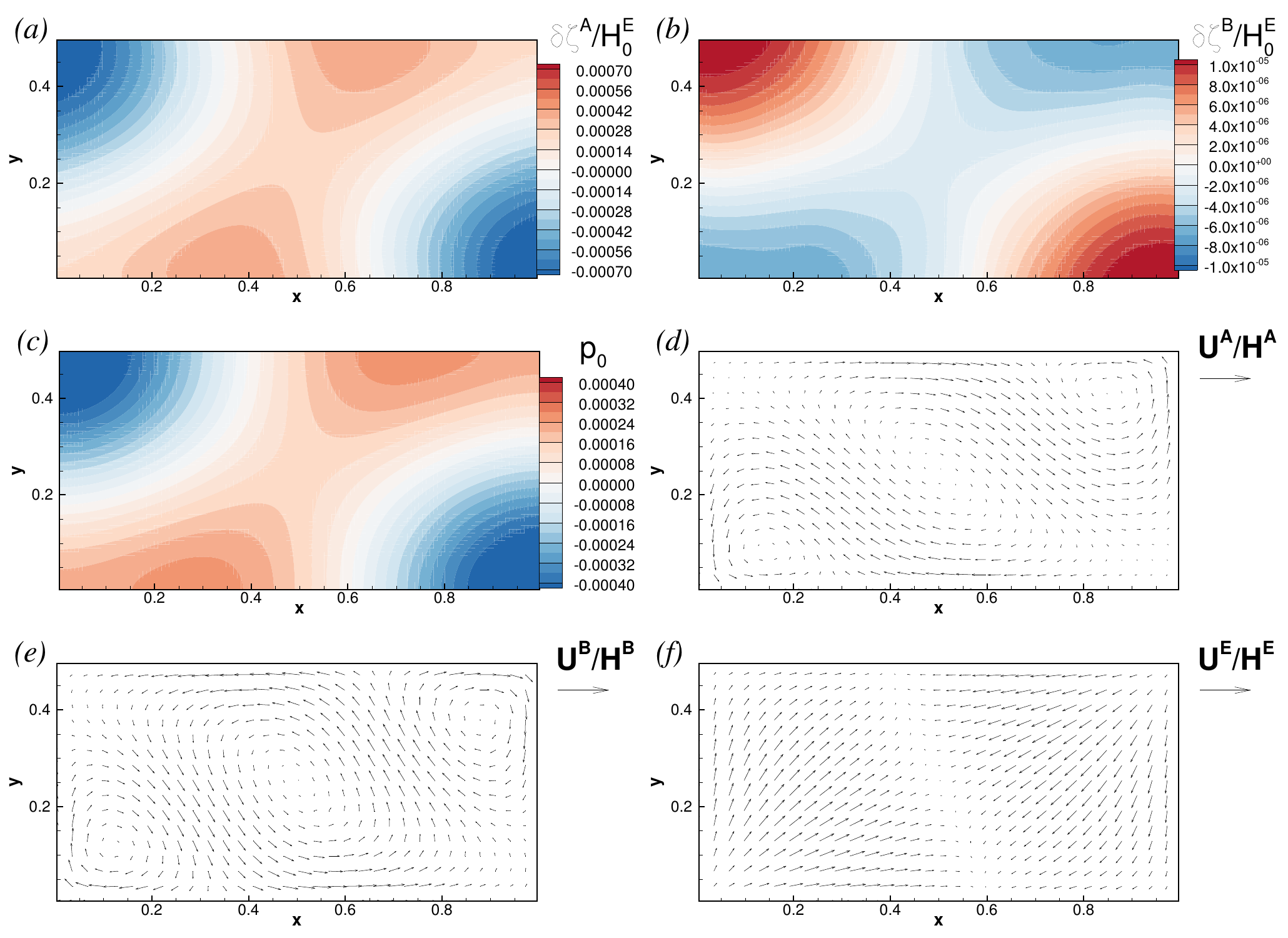}
\caption{Example of the rolling pad instability in the case when only the upper interface is significantly deformed. The flow parameters are as in Fig.~\ref{fig:singleoutput}. Instantaneous distributions of the deformations of upper \emph{(a)} and lower  \emph{(b)} interfaces, mid-plane pressure \emph{(c)}, and vertically {averaged} velocities $\bm{U}^A/H^A$,  $\bm{U}^B/H^B$,  $\bm{U}^E/H^E$  \emph{(d)-(f)} are shown at $t$=31.1 (see Fig.~\ref{fig:singleoutput}). The interface deformations are scaled by the unperturbed electrolyte thickness $H_0^E$. Every fourth vector in each direction is plotted in \emph{(d)-(f)}. In each picture, vector's length is proportional to the local velocity amplitudes. {The reference vectors correspond to velocity amplitude of $10^{-3}$, $10^{-4}$, and $5\times 10^{-3}$ in, respectively, \emph{(d)}, \emph{(e)}  and \emph{(f)}.}   }
\label{fig:singlestructure}
\end{center}
\end{figure}

As a typical example of the instability observed at $\Delta \rho^A\ll \Delta \rho^B$, Figs.~\ref{fig:singleoutput} and \ref{fig:singlestructure} show the flow in a rectangular cell with the aspect ratio $\Gamma=2.0$, $L_x=0.75$ m, $\rho^A=1000$ kg/m$^3$, $\rho^B=8000$ kg/m$^3$, $\rho^E=1100$ kg/m$^3$, $J_0=10^4$ A/m$^2$, $B_0$=0.001 T, $H_0^A=H_0^B=0.1$ m, $H_0^E=0.005$ m, and $\nu^A=\nu^B=\nu^E=5\times 10^{-7}$ m$^2$/s.  

After an initial adjustment, the time signals of the local and integral characteristics show combined periodicity and exponential growth. Analysis of the interface deformations such as those in Figs.~\ref{fig:singlestructure}a,b shows that these features are caused by, respectively, rotation and growth of an interfacial wave. At the positive (directed upward) base current $J_0$, the rotation is counterclockwise if viewed from above. A negative base current changes the sense of rotation, but does not affect the flow in any other way. 

The solution in Figs.~\ref{fig:singleoutput} and \ref{fig:singlestructure} is qualitatively similar to the solutions obtained in \cite{Weber:2016,Bojarevics:2017} in terms of the spatial shape, dynamics, and typical length and time scales of the growing perturbations. The amplitude of the deformation is much smaller at the lower than at the upper interface, approximately in proportion to $\Delta \rho^A/\Delta \rho^B$. 

The instability can be characterized by two coefficients computed during the phase of exponential growth: the time period of oscillations $T$ and the growth rate $\gamma$. Both can be estimated using the  time signals shown in Fig.~\ref{fig:singleoutput}. For example, the point signal of the interface deformation in Fig.~\ref{fig:singleoutput}a can be used to determine $T$, while an exponential fit of the appropriate segments of the curves in Fig.~\ref{fig:singleoutput}b,c provides $\gamma$. For the specific flow in Fig.~\ref{fig:singleoutput}, the values are $T=0.947$ and $\gamma=0.379$.

The unstable wave illustrated in Figs.~\ref{fig:singleoutput} and \ref{fig:singlestructure} grows until the maximum interface deformation exceeds $0.5H_0^E$, at which point the simulations are stopped. Such a growth is typically observed in our simulations in strongly unstable systems. In systems with positive but small growth rate $\gamma$, the perturbations may saturate nonlinearly at some finite amplitude of the interface deformation, typically six to three times smaller than $H_0^E$. 

The evident similarity with the instability in the aluminum reduction cells indicates that the observed instability may be controlled by the same non-dimensional parameters:  the horizontal aspect ratio $\Gamma$,  the instability parameter, which we discuss shortly, and the Reynolds number, which appears since viscosity is included in our model. 

An expression for the parameter controlling the instability is suggested by the studies of the reduction cells. The simplified models, in which viscosity and other complicated effects are neglected, and modified wave equations for low-amplitude large-scale perturbations are derived, all produce principally the same results (see \cite{Sele:1977,Urata:1985,Sneyd:1994,Bojarevics:1994,Davidson:1998}). The parameter is directly proportional to the product $J_0B_0$ and the square of the horizontal dimension of the cell, and inversely proportional to the product of the thicknesses of the metal and electrolyte layers, density difference $\Delta \rho$, and gravity acceleration. In our notation, for the instability developing at the upper interface, this becomes
\begin{equation}
\label{pi1}
\Pi \equiv \frac{J_0B_0L_xL_y}{\Delta \rho^A H_0^EH_0^Ag}.
\end{equation}
A similar parameter with $\Delta \rho^B$ instead of $\Delta \rho^A$ should be used at $\Delta \rho^B\ll \Delta \rho^A$ when the instability develops at the lower interface. 

The expression (\ref{pi1}) is equivalent to the parameter $\beta$ derived for the reduction cells in the pioneering work of \cite{Sele:1977} and to the parameter, in terms of which the results of \cite{Davidson:1998} were recently recast in \cite{Davidson:2016}. It can be rewritten in terms of the non-dimensional parameters introduced in section \ref{sec:nondim} as
\begin{equation}
\label{pi1nd}
\Pi=\epsilon \Fr^2 \Gamma \frac{1}{\tilde{H}_0^A \tilde{H}_0^E}\frac{1}{\gamma_\rho^A-1}.
\end{equation}
  
We have carried out  extensive testing in order to verify that $\Pi$ is a suitable control parameter for the single-interface instability in the battery. Firstly, we verified that a significant change of $\Pi$ leads to a significant change of the wave characteristics $\gamma$ and $T$.  Furthermore, tests were performed, in which the  battery parameters, such as $L_x$, $B_0$, $J_0$, $H_0^A$, or $H_0^E$ were varied substantially in such a way that the value of $\Pi$ remained the same. The results of one of such tests conducted at $\Gamma=2.0$ are shown in table \ref{tab:roll}. Two sets of simulations are exhibited: at the normal values of viscosities and at the values reduced by four orders of magnitude, which practically removed the effect of viscosity on the flow. 

We see in table  \ref{tab:roll} that at negligible viscosity the characteristics of the growing wave $\tilde{\gamma}$ and $\tilde{T}$ are determined by $\Pi$ with very good accuracy. The computed values are within 1\%. An exception is the simulation with $\Delta \rho^A$ increased to 200 kg/m$^3$, in which $\tilde{\gamma}$ and $\tilde{T}$ change by about 3\%. This effect will be further discussed in section \ref{sec:paddouble}.

The variation is stronger at the realistic $\nu=5\times 10^{-7}$ m$^2$/s. This can be attributed to the non-negligible effect of the variation of the Reynolds number.

\begin{table}
  \begin{center}
  \begin{tabular}{c|c|c|c|c|c|c}
  Run & Base & $J_0=5\times 10^3$  &   $J_0=5\times 10^3$  & $J_0=2\times 10^4$  & $J_0=2\times 10^4$  & $J_0=2\times 10^4$ \\
        &   case       & $L_x=1.061$            &   $B_0=0.002$               & $H_0^E=0.01$        & $H_0^A=H_0^B=0.2$            & $\Delta \rho^A=200$ \\
       \hline
 $\gamma$  & 0.379 & 0.357 & 0.378 & 0.417 & 0.404 & 0.384 \\
  $ T$ & 0.947 & 0.946 & 0.947 & 0.945 & 0.947 & 0.952 \\
       \hline
 $\tilde{\gamma}$  & 0.439 & 0.439 & 0.439 & 0.440 & 0.438 & 0.425 \\
 $ \tilde{T}$ & 0.944 & 0.943 & 0.944 & 0.948 & 0.946 & 0.952
           \end{tabular}
  \end{center}
  \caption{Verification test of the instability parameter (\ref{pi1}). Exponential growth rates $\gamma$ and periods $T$ of growing perturbations at viscosity $\nu^A=\nu^B=\nu^E=5\times 10^{-7}$ m$^2$/s and the analogous coefficients $\tilde{\gamma}$, $ \tilde{T}$ at $\nu^A=\nu^B=\nu^E=5\times 10^{-11}$ m$^2$/s are shown. The base case is the simulation with $\Pi=5.734$ illustrated in Figs.~\ref{fig:singleoutput} and \ref{fig:singlestructure}. In each of the other  5 simulations, two battery parameters are changed in comparison to the base case so that the value of $\Pi$ remains the same.  Current density $J_0$ is in A/m$^2$, magnetic field $B_z$ is in T, lengths are in m, and density is in kg/m$^3$.}
  \label{tab:roll}
\end{table}

We have also performed simulations with $\Delta \rho^B\ll \Delta \rho^A$, i.e. when the significant  deformations are expected only on the lower interface. The results are qualitatively the same as when $\Delta \rho^A\ll \Delta \rho^B$. The only significant differences are the opposite sense of rotation (clockwise at $J_0>0$ and counterclockwise otherwise), which is evidently caused by the opposite direction of the horizontal current perturbations in the bottom metal layer (see (\ref{chargecons}) and the illustration in Fig.~\ref{fig1}b), and the change of $T$ and $\gamma$ attributed to the change of the typical time scale.  

\begin{figure}
\begin{center}
\includegraphics[width=1.0\textwidth]{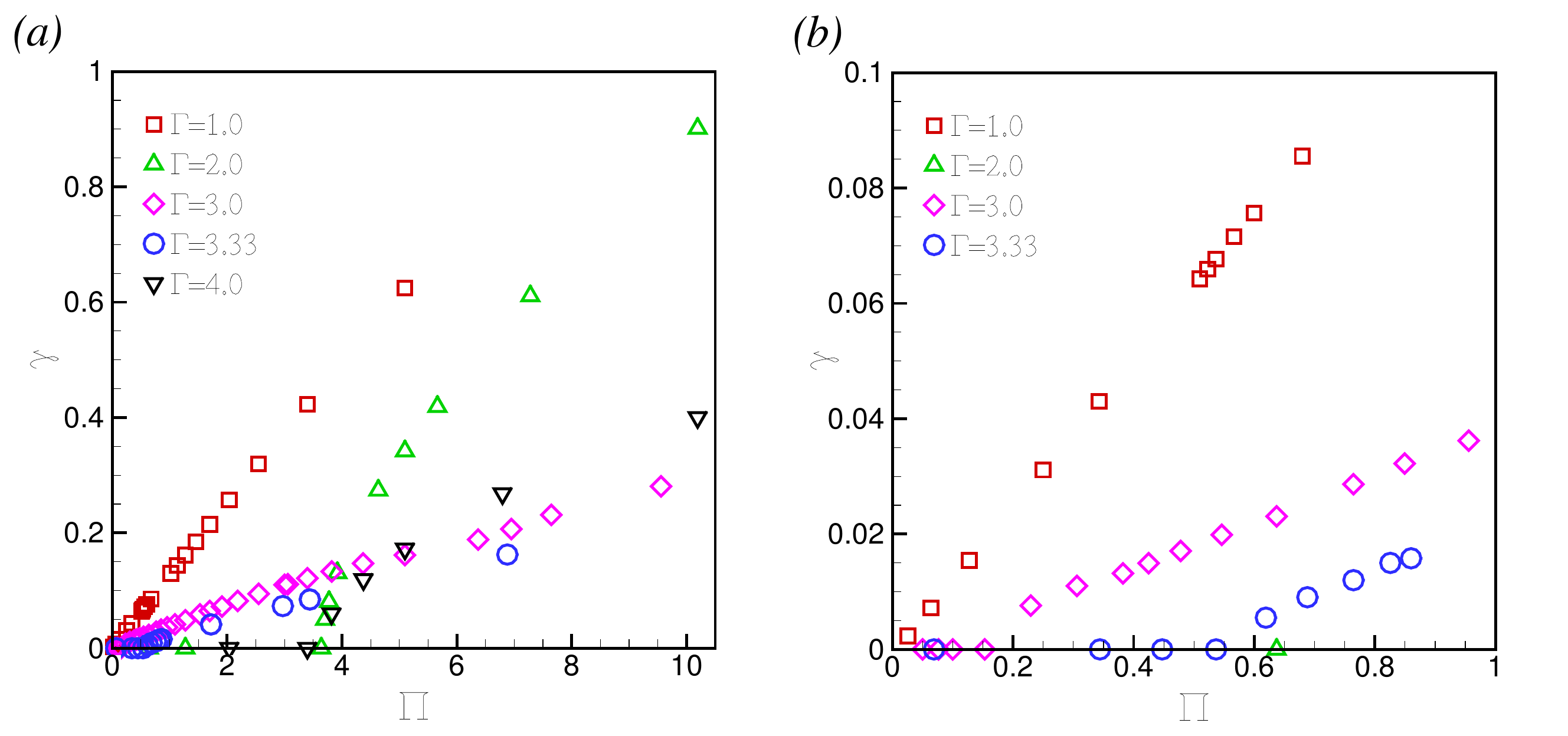}
\caption{Parametric study of the rolling pad instability in the case when only the upper interface is significantly deformed. The exponential growth rate $\gamma$ is shown as a function of the instability parameter $\Pi$ (see (\ref{pi1})) for cells of various horizontal aspect ratios $\Gamma$ with $\Delta \rho^A=100$ kg/m$^2$.  $\gamma=0$ is used for the cases where the battery is stable. The plot \emph{(b)} is a zoom-in to small values of $\Pi$.   }
\label{fig:singleresults}
\end{center}
\end{figure}

Parametric studies were performed to explore the effect of $\Pi$ and the aspect ratio $\Gamma$ on the instability. The realistic values $\nu^A=\nu^B=\nu^E=5\times 10^{-7}$ m$^2$/s were used. In the study, the same values $\rho^A=1000$ kg/m$^3$, $\rho^E=1100$ kg/m$^3$, $\rho^B=8000$ kg/m$^3$, and $H_0^A=H_0^B=0.1$ m were used, while the other parameters were {varied in the ranges $J_0=5\times 10^3$, $10^4$, A/m$^2$ $B_0=10^{-3}$, $2\times 10^{-3}$ T, $0.5\le L_x\le 1.5$ m, and $0.005\le H_0^E\le 0.1$ m so as to change the value of $\Pi$.} The results are presented in Fig.~\ref{fig:singleresults}. We find that, as one would expect by analogy with the reduction cells, the behavior is strongly influenced by $\Gamma$. In particular, the instability threshold $\Pi_{cr}$ such that the battery is stable at all $\Pi<\Pi_{cr}$ but unstable at $\Pi>\Pi_{cr}$, is a non-monotonic function of $\Gamma$. At $\Gamma=1.0$, $\Pi_{cr}$ is very close to zero, although it cannot be exactly zero due to the stabilizing effect of viscosity and numerical dissipation. In rectangular cells, $\Pi_{cr}$ can be small, as, for example, $\Pi_{cr}(3.0)\approx 0.15$ or large as $\Pi_{cr}(2.0)\approx 3.7$ or $\Pi_{cr}(4.0)\approx 3.5$.  Even small variations of $\Gamma$ may change $\Pi_{cr}$ quite significantly. An example is the difference between $\Pi_{cr}(3.0)\approx 0.15$ and $\Pi_{cr}(3.33)\approx 0.67$. This and the strong effect of $\Gamma$ in general can be explained in the same way as, \eg in \cite{Davidson:1998} for the aluminum reduction cells. The aspect ratio determines the set of available natural gravitational wave modes and, so, strength of the electromagnetic effect needed to transform two of them into a pair with complex-conjugate eigenvalues.

\begin{figure}
\begin{center}
\includegraphics[width=0.65\textwidth]{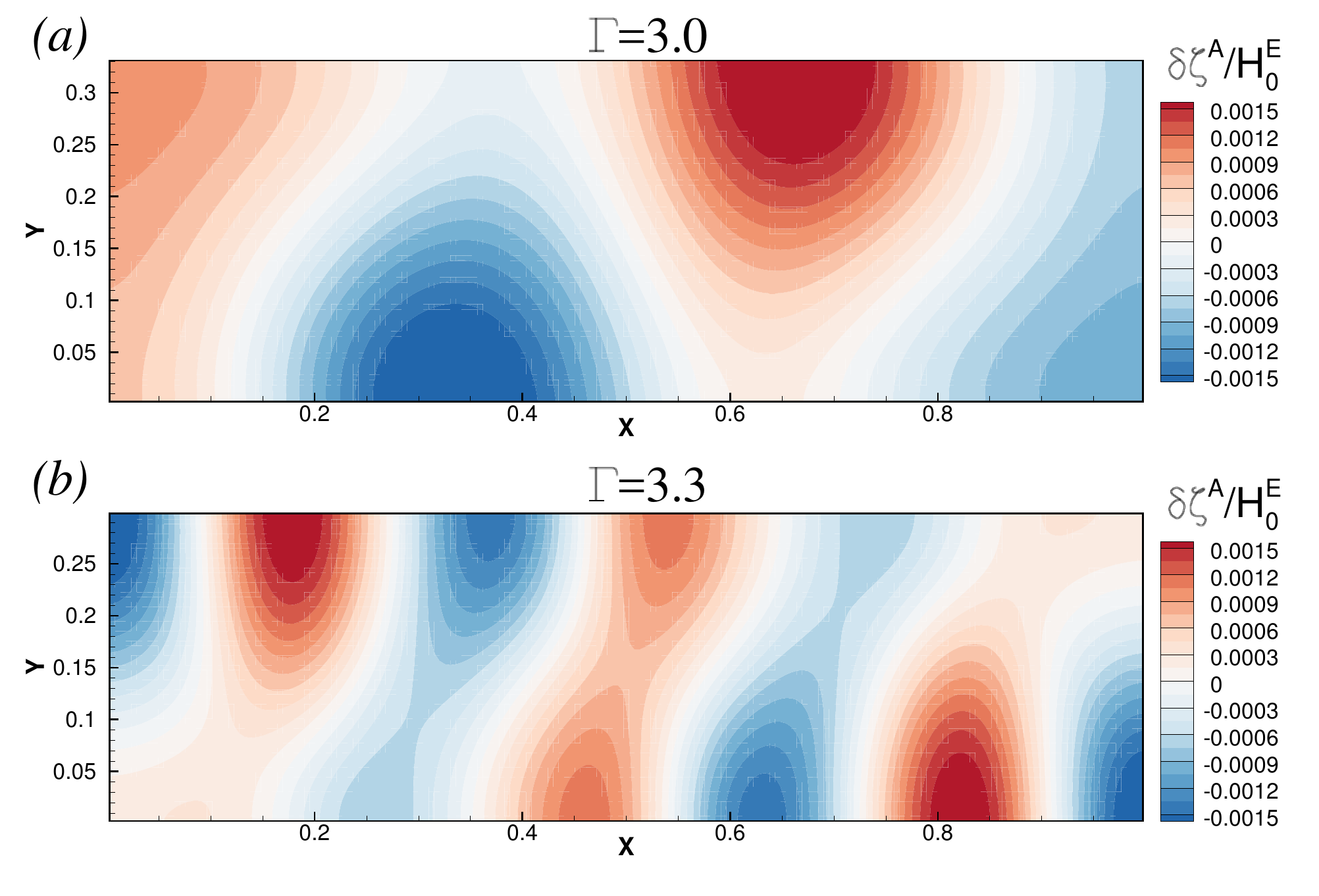}
\caption{The effect of the aspect ratio $\Gamma$ on unstable wave in the case when only the upper interface is significantly deformed. Instantaneous deformations of the  interface at $\Pi=3.0$, and $\Gamma=3.0$ \emph{(a)} and 3.33 \emph{(b)} are shown (see text for other flow parameters).  The interface deformations are scaled by the unperturbed electrolyte thickness $H_0^E$.  The moments of the flow evolution are selected for both the cases within the  exponential growth range and so that the amplitudes of the deformation are approximately equal.  }
\label{fig:aspect}
\end{center}
\end{figure}

As an illustration of the aspect ratio effect, Fig.~\ref{fig:aspect} shows the interface deformations computed for the unstable flows at $\Gamma=3.0$ (Fig.~\ref{fig:aspect}a) and $\Gamma=3.33$ (Fig.~\ref{fig:aspect}a). Almost all the battery parameters are the same in the two cases: $J_0=10^4$ A/m$^2$, $B_0=0.001$ T, $L_x=1.5$ m, $\rho^A=1000$ kg/m$^3$, $\rho^E=1100$ kg/m$^3$, $\rho^B=8000$ kg/m$^3$, $H_0^A=H_0^B=0.1$ m. The only differences are in the values of $L_y$ (0.5 m or 0.45 m) and the values of $H_0^E$ (0.02548 m or 0.02317 m) that are selected in such a way that the instability parameter is $\Pi=3.0$ in both cases. The time signals of the two unstable solutions show strongly different characteristics: $T=0.690$, $\gamma=0.110$ at $\Gamma=3.0$ and $T=0.335$, $\gamma=0.073$ at $\Gamma=3.33$. The growing interface deformations shown in Fig.~\ref{fig:aspect} demonstrate that the explanation suggested above is valid. The electromagnetic coupling, which causes the instability, involves different gravitational wave modes in the two cases, with much larger $x$-wavelength at $\Gamma=3.0$ than at  $\Gamma=3.33$. A more specific description can be obtained with the eigenvalue analysis.  

The main conclusion of this section is that the single-interface rolling pad instability occurring when one density difference is much smaller than the other is very similar to the instability observed in the aluminum reduction cells. The instability is caused by the same mechanism, develops in essentially the same form of a rotating interfacial wave, and is determined by three non-dimensional parameters: the instability parameter (\ref{pi1}) (or its analog based on $\Delta \rho^B$, $H_0^B$ instead of $\Delta \rho^A$, $H_0^A$ if the instability occurs on the lower interface), the Reynolds number $\Rey$, and the horizontal aspect ratio $\Gamma$.


\subsection{Double interface deformation}\label{sec:paddouble}
The situation when $\Delta\rho^A$ and $\Delta\rho^B$ are of comparable magnitudes, and, thus, the instability may significantly deform both the interfaces, is considered in this section. The analysis is performed as a parametric study, in which the metal densities are kept constant at $\rho^A=1000$ kg/m$^3$ and $\rho^B=8000$ kg/m$^3$, while the density of the electrolyte $\rho^E$ varies in the range between $1100$  and $6000$ kg/m$^3$. This can be compared with the material densities in the currently considered battery schemes (see \cite{Kim:2013}): 500 to 1500 kg/m$^3$ for $\rho^A$, 6000 to 10000 kg/m$^3$ for $\rho^B$, and 1000 to 3000 kg/m$^3$ for $\rho^E$. {In the study, we keep constant metal layer thicknesses $H^A=H^B=0.1$ m and viscosities $\nu^A=\nu^B=\nu^E=5\times 10^{-7}$ m$^2$/s. The other parameters vary as $5\times 10^3\le J_0\le 4\times 10^4$ A/m$^2$, $10^{-3}\le B_0\le 3\times 10^{-2}$ T, $1\le L_x\le 2$ m, $0.005 \le H^E\le 0.04$ m. The behavior at the aspect ratios $\Gamma=2.0$ and 3.0 is explored.}

\begin{figure}
\begin{center}
\includegraphics[width=1.0\textwidth]{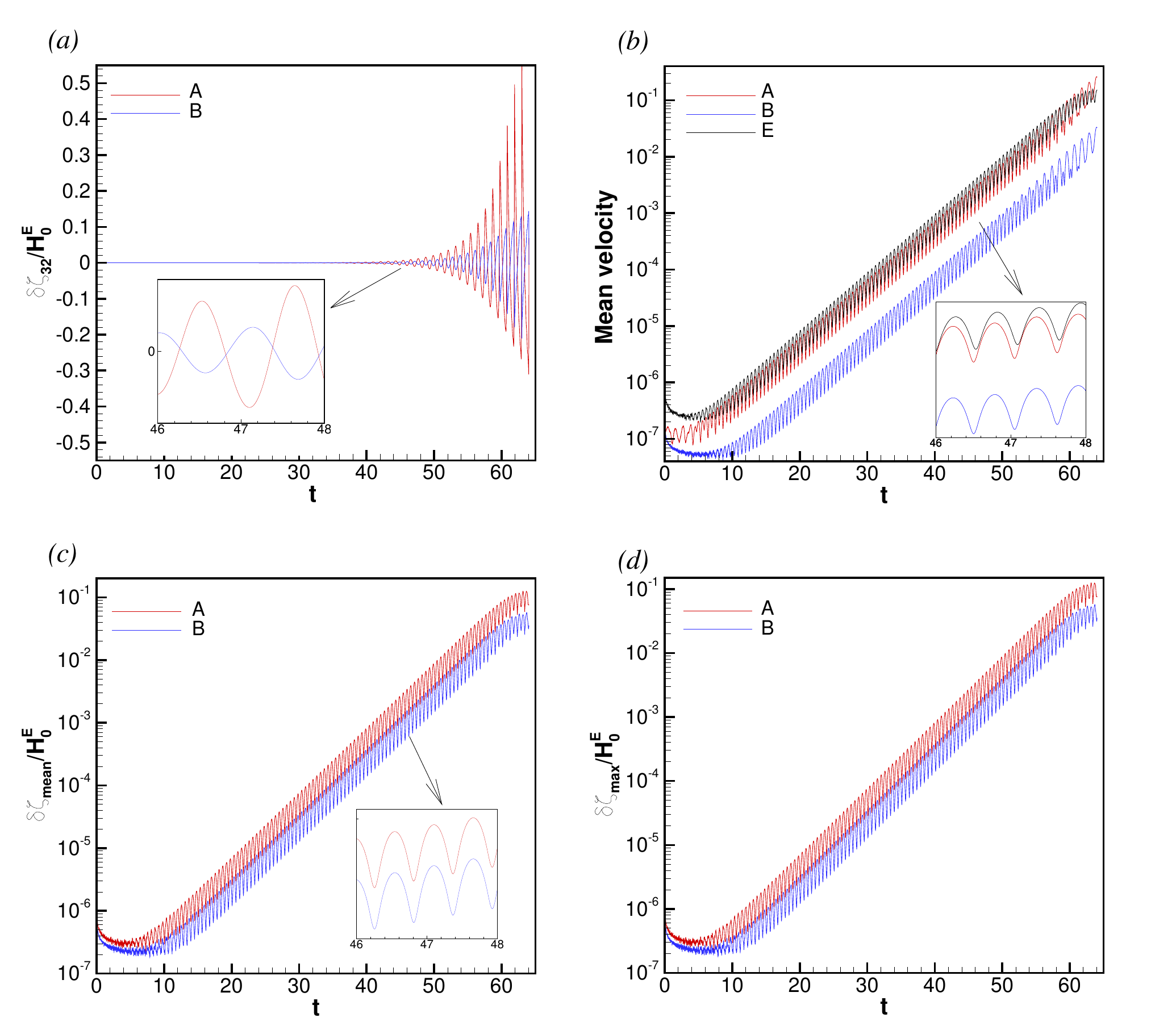}
\caption{Example of the rolling pad instability in the case when both the interfaces are significantly deformed (see text for system's parameters). Deformation of the two interfaces the grid point $(x_3,y_2)$ \emph{(a)}, space-averaged rms of fluid velocities $U/H$ in each layer \emph{(b)}, space-averaged rms of interface deformations \emph{(c)}, and maximum interface deformations \emph{(d)} are shown as functions of time for the entire simulated flow evolution. The interface deformations are scaled by the unperturbed electrolyte thickness $H_0^E$. }
\label{fig:doubleoutput}
\end{center}
\end{figure}

\begin{figure}
\begin{center}
\includegraphics[width=1.0\textwidth]{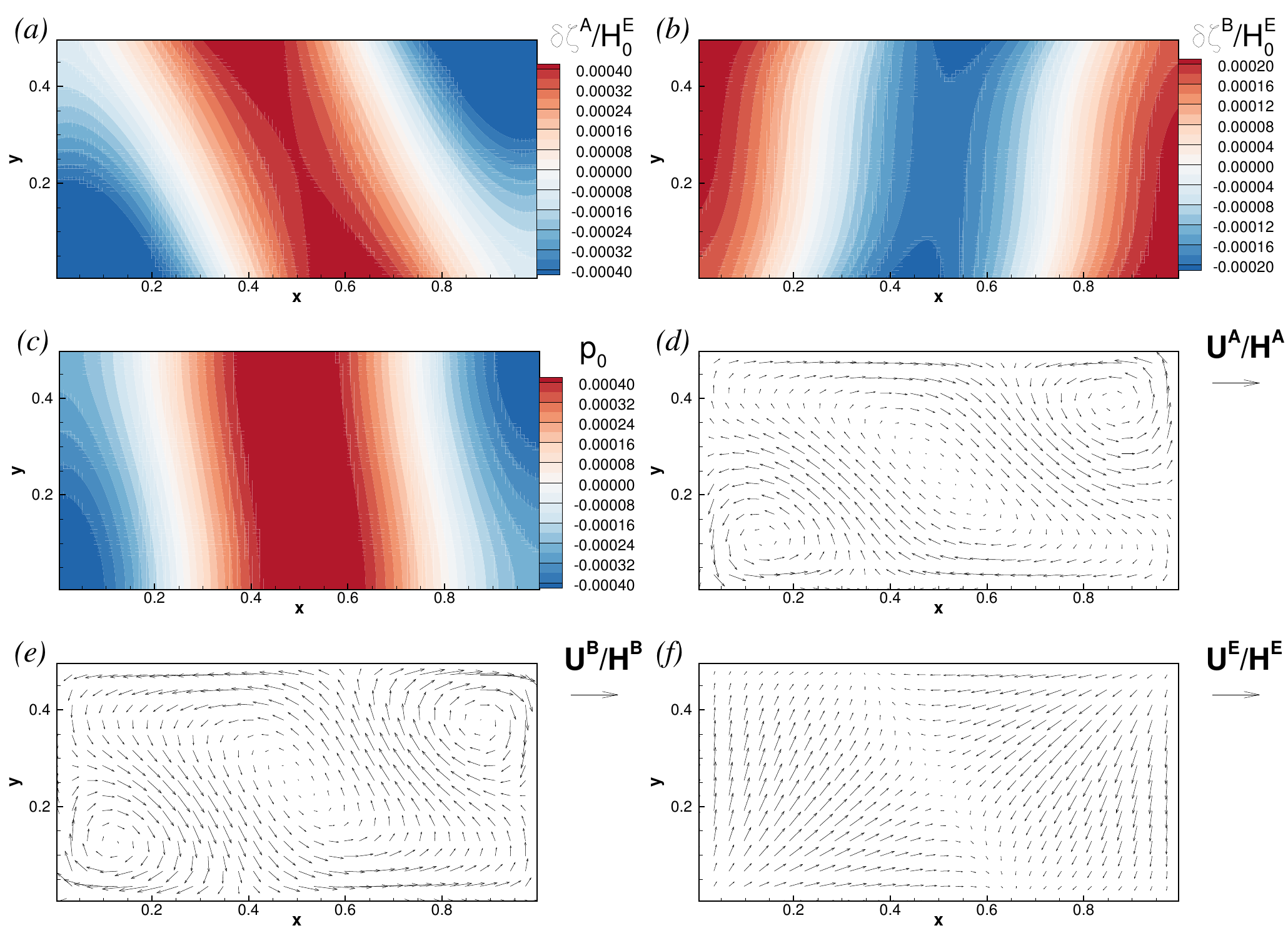}
\caption{Example of the rolling pad instability in the case when both the interfaces are significantly deformed. The flow parameters are as in Fig.~\ref{fig:doubleoutput}. Instantaneous distributions of the deformations of upper \emph{(a)} and lower  \emph{(b)} interfaces, mid-plane pressure \emph{(c)}, and vertically {averaged} velocities $\bm{U}^A$,  $\bm{U}^B$,  $\bm{U}^E$  \emph{(d)-(f)} are shown at $t$=38.0 (see Fig.~\ref{fig:doubleoutput}). The interface deformations are scaled by the unperturbed electrolyte thickness $H_0^E$.  Every fourth vector in each direction is plotted in \emph{(d)-(f)}. In each picture, vector's length is proportional to the local velocity amplitudes. {The reference vectors correspond to velocity amplitude of $2.5\times 10^{-3}$, $2.5\time 10^{-4}$, and $4\times 10^{-3}$ in, respectively, \emph{(d)}, \emph{(e)}  and \emph{(f)}.}   }
\label{fig:doublestructure}
\end{center}
\end{figure}

\begin{figure}
\begin{center}
\includegraphics[width=1.0\textwidth]{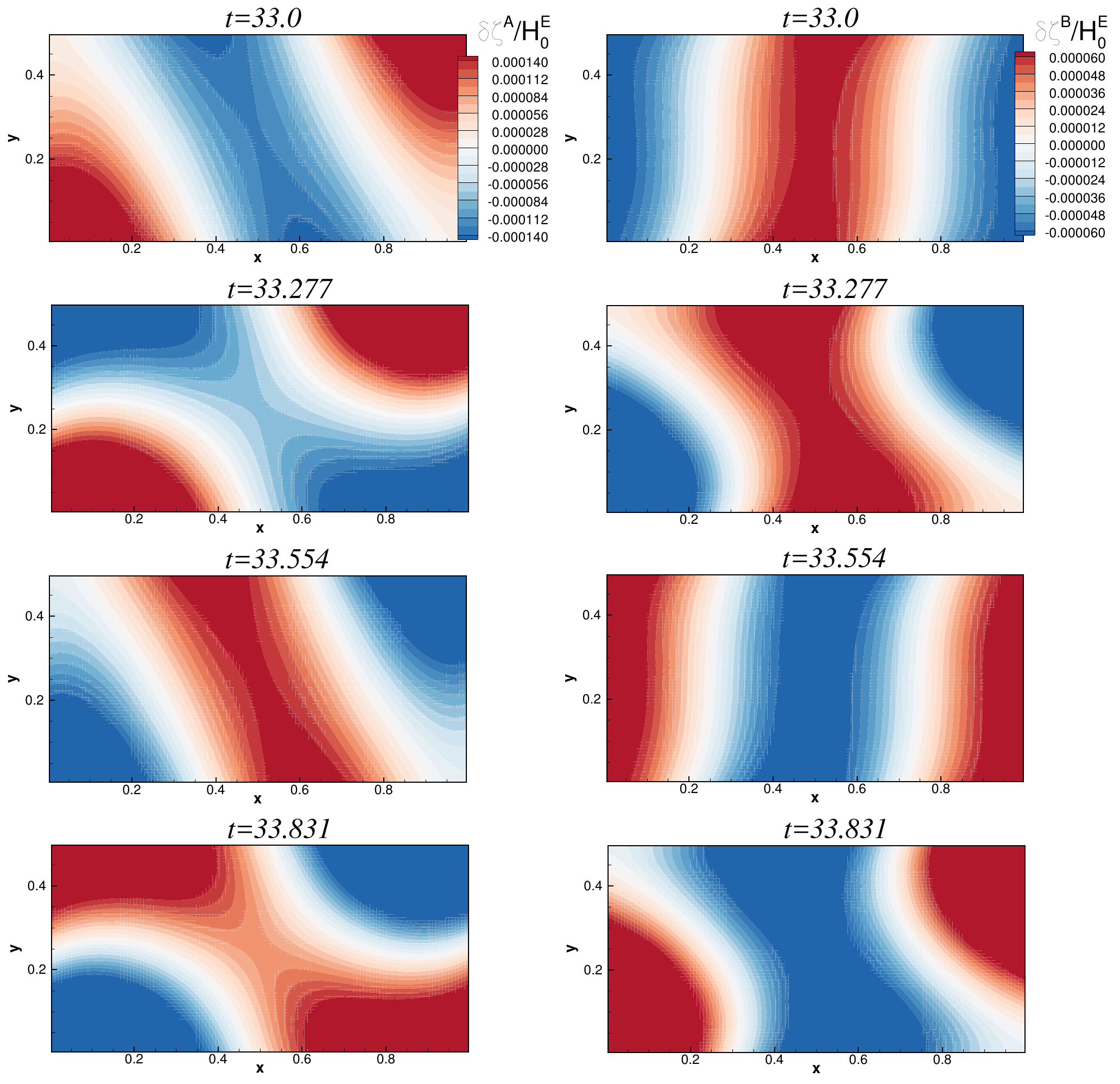}
\caption{Example of the rolling pad instability in the case when both the interfaces are significantly deformed. The flow parameters are as in Figs.~\ref{fig:doubleoutput} and \ref{fig:doublestructure}. Snapshots of the deformations of the upper (left column) and lower (right column) interfaces are shown for four time moments taken every quarter of the oscillation period.   }
\label{fig:doubleperiod}
\end{center}
\end{figure}

\begin{figure}
\begin{center}
\includegraphics[width=0.50\textwidth]{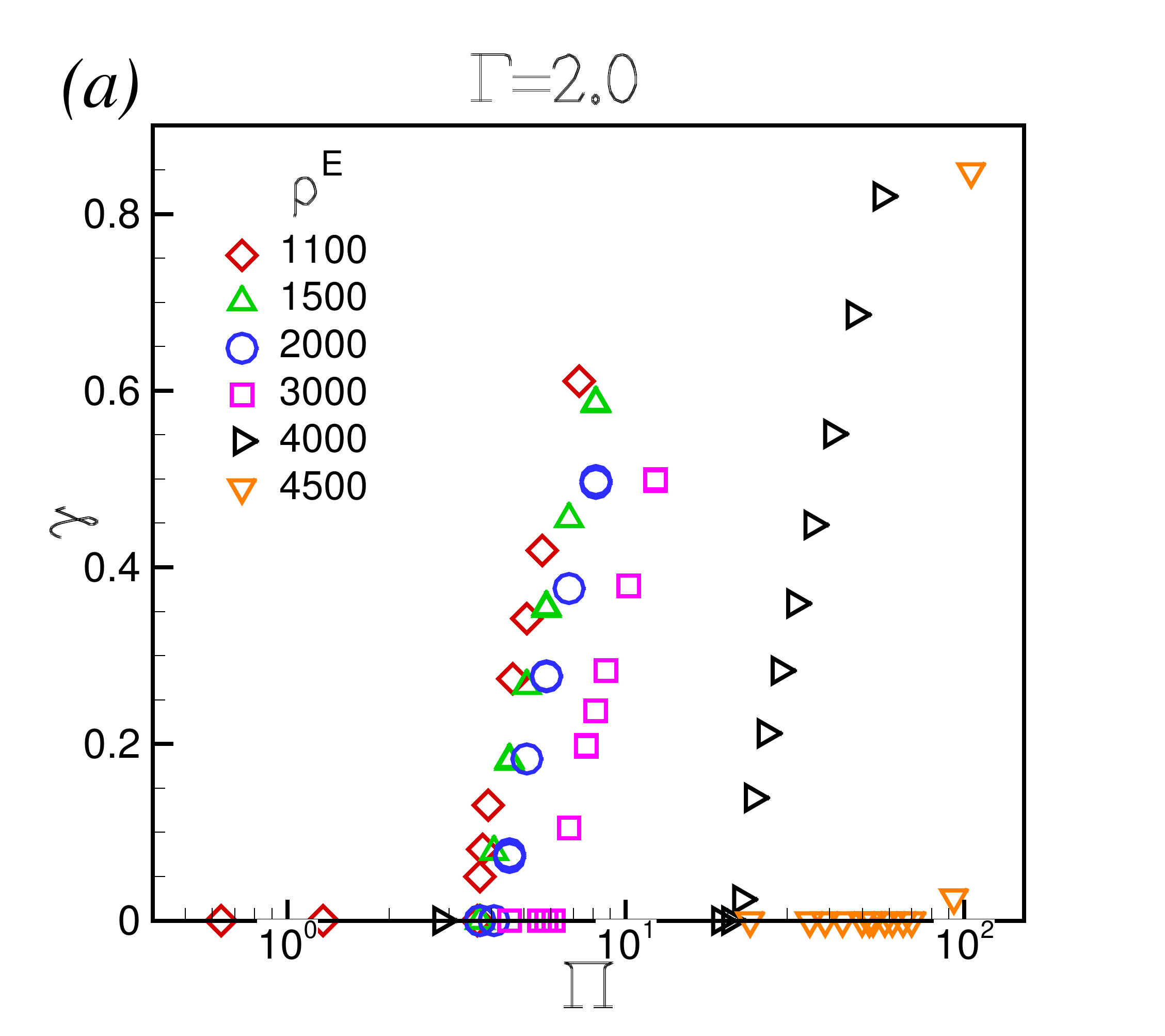}\includegraphics[width=0.50\textwidth]{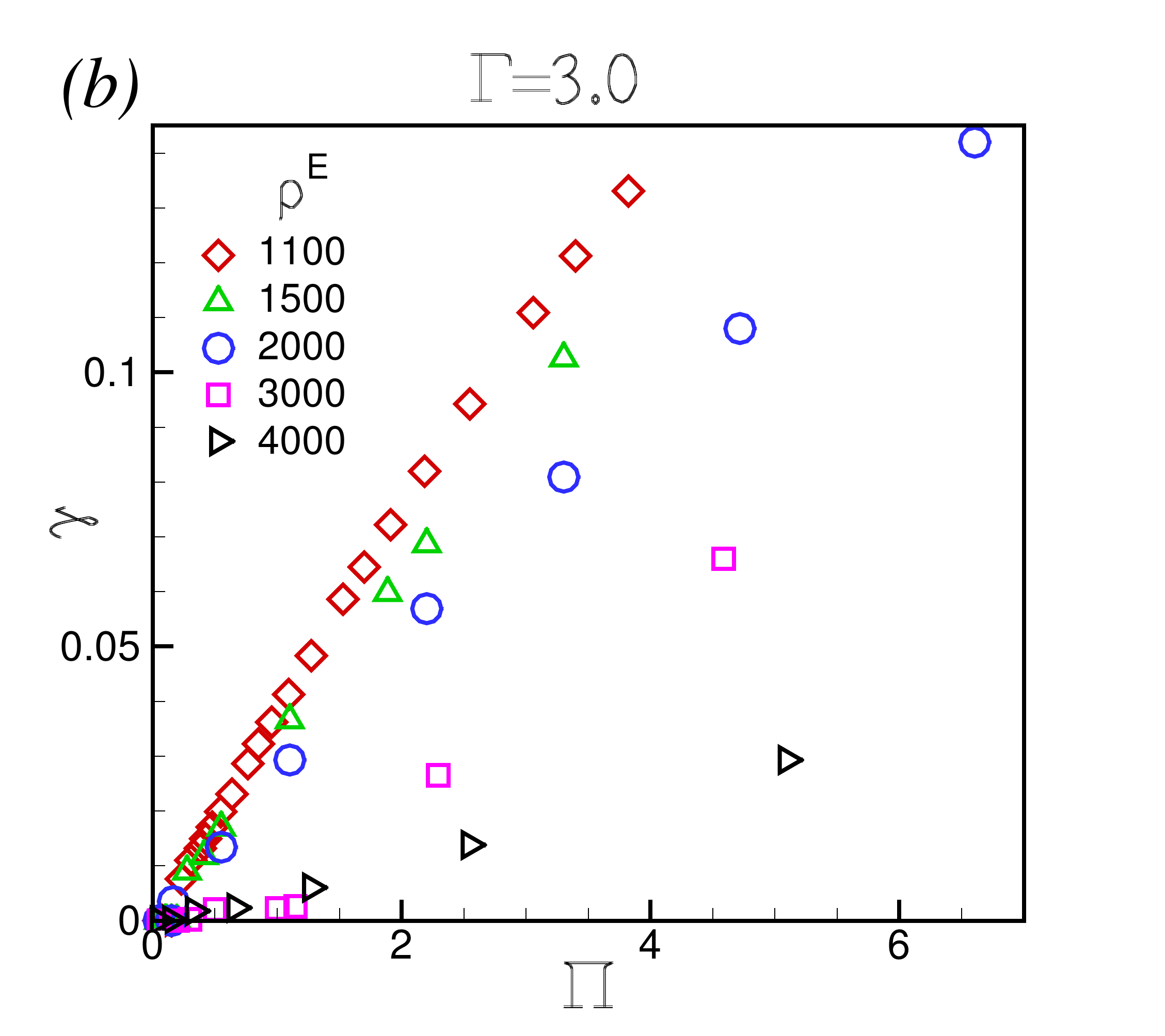}
\caption{Parametric study of the rolling pad instability in the case when both the interfaces are significantly deformed. The exponential growth rate $\gamma$ is shown as a function of the non-dimensional parameter $\Pi$ (see (\ref{pi1})) for cells with $\Gamma=2.0$ \emph{(a)} and 3.0 \emph{(b)}, $\rho^A=1000$ kg/m$^3$, $\rho^B=8000$ kg/m$^3$ and various values of $\rho^E$. }
\label{fig:doubleresults}
\end{center}
\end{figure}

As a typical example, Figs.~\ref{fig:doubleoutput}-\ref{fig:doubleperiod} show the instability in a system with $\rho^E=3000$ kg/m$^3$, $J_0=2\times 10^4$ A/m$^2$, $B_0=0.002$ T, $L_x=2.0$ m, $L_y=1.0$ m, $H_0^A=H_0^B=0.1$ m, $H_0^E=0.005$ m. The results of the parametric studies are summarized in Fig.~\ref{fig:doubleresults}.

We see in Figs.~\ref{fig:doubleoutput}-\ref{fig:doubleperiod}  that the main features of the instability remain the same as in the case of the single-interface deformation. The time signals combine periodic oscillations with exponential growth. The spatial structure in Fig.~\ref{fig:doublestructure} shows large-scale interfacial waves quite similar to the wave in Fig.~\ref{fig:singlestructure}. A conclusion can be made that the instability is still of the rolling pad type. 

\begin{figure}
\begin{center}
\includegraphics[width=1.0\textwidth]{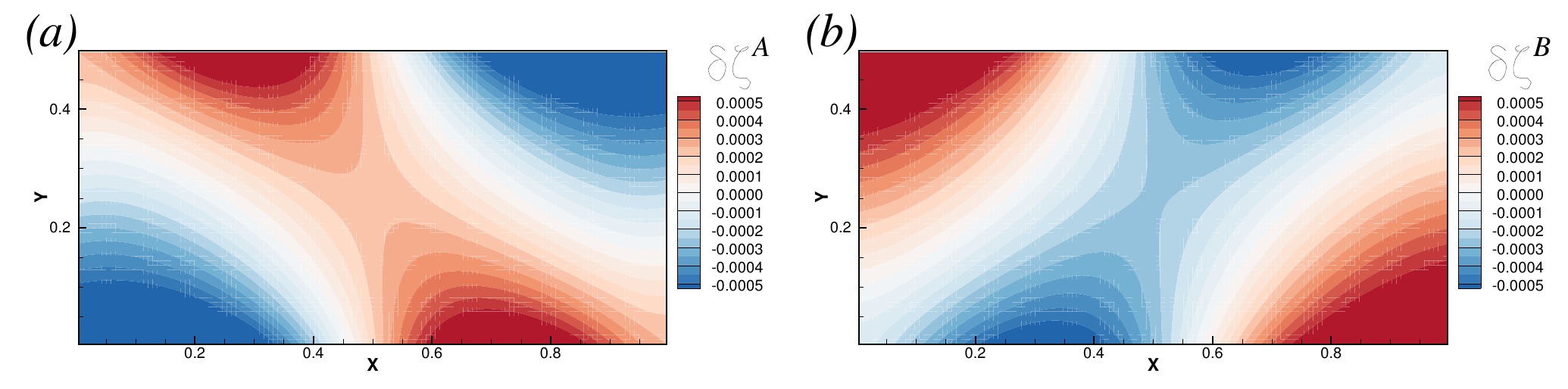}
\caption{Example of growing nearly symmetric  waves at $\Delta \rho^A \sim \Delta \rho^B$. Interface deformations  during the stage of exponential growth are shown for the system with $\rho^A= 1000$ kg/m$^3$, $\rho^E= 4000$ kg/m$^3$, $\rho^B= 8000$ kg/m$^3$, $J_0=3\times 10^4$ A/m$^2$, $B_0=0.007$ T, $L_x=2.0$ m, $L_y=1.0$ m, $H_0^A=H_0^B=0.1$ m, $H_0^E=0.005$ m. }
\label{fig:symmetric}
\end{center}
\end{figure}

At the same time, the interaction between the two waves growing at the lower and upper interfaces results in more complex and diverse dynamics. The waves are always coupled to each other in the sense that they have the same oscillation period and sense of rotation (counterclockwise at $\Delta \rho^A<\Delta \rho^B$ and clockwise at $\Delta \rho^A>\Delta \rho^B$ if $J_0>0$). Two types of coupling are observed. At substantially different $\Delta \rho^A$ and $\Delta \rho^B$ (\eg at $\rho^E\le 3000$ kg/m$^3$ or $\rho^E\ge 6000$ kg/m$^3$ at $\Gamma=2.0$), the waves at the upper and lower interfaces are nearly antisymmetric (the sign of $\delta \zeta^A$ is opposite to the sign of $\delta \zeta^B$). The antisymmetry is not exact, but with a small phase shift. For example, the time signals of the single-point oscillations in Fig.~\ref{fig:doubleoutput}a show the time shift $\Delta t\approx 0.042$, which can be compared with the oscillation period of 1.114. The situation changes at close $\Delta \rho^A$ and $\Delta \rho^B$. In such cases, the two waves become nearly symmetric (with the signs of $\delta \zeta^A$ and $\delta \zeta^B$ being mostly the same) and, again, with a small phase shift. This type of behavior is illustrated in Fig.~\ref{fig:symmetric}.

{The observed change of the type of wave coupling is consistent with the results of the  recent three-dimensional potential flow analysis of interfacial waves in a three-layer cylindrical cell without electromagnetic forces \cite{Horstmann:2017}. It has been shown in this work that  the type of coupling between the waves at the two interfaces changes from antisymmetric to symmetric when the amplitudes of the waves become comparable. While not a proof due to the evident differences between the systems, the consistency can be considered as an indication that the change of coupling observed in our battery is likely to be related to purely hydrodynamic interaction between the waves, rather than to electromagnetic forces. }

The results of the parametric study presented in Fig.~\ref{fig:doubleresults} show that the effect of the relation between $\Delta \rho^A$ and $\Delta \rho^B$ on stability is not very significant in the strongly unstable system with $\Gamma=3.0$. The values of the instability growth rate $\gamma$ decrease as the two density differences become closer to each other, but the instability continues to occur at small values of $\Pi$.  

On the contrary, for the system with $\Gamma=2.0$, increase of  $\Delta \rho^A$ has stabilizing effect, which is not only substantial, but also clearly exceeding the effect of the density stratification already incorporated into the expression (\ref{pi1}) of the instability parameter. The critical values $\Pi_{cr}$, such that the system is always stable at $\Pi<\Pi_{cr}$ but unstable at $\Pi>\Pi_{cr}$, can be estimated as $\Pi_{cr}=3.7$, 4.0, 4.5, 6.2, and 21.5 at $\Delta \rho^A=100$, 500, 1000, 2000, and 3000 kg/m$^3$, respectively. This growth is in contradiction with the prediction of the mechanical model \cite{Zikanov:2015}, where the presence of the second pendulum was found to be always destabilizing. We attribute the contradiction to the oversimplified character of the model used in \cite{Zikanov:2015}. 

The stabilizing effect of the second layer can be given a qualitative explanation based on the fact that the horizontal perturbation currents in the top and bottom layers are always flowing in the opposite directions (see (\ref{chargecons})). This implies the opposite directions of the horizontal Lorentz forces (\ref{ab-forces}) and, thus, that forces in one layer always oppose the motion of a coupled wave system. Whether this happens in the bottom or top layer is determined by the relation between $\Delta \rho^A$, $\Delta \rho^B$, $H_0^A$, $H_0^B$ and, possibly, other parameters. 


\section{Concluding remarks}\label{sec:conclusion}
We have presented a new shallow water model and the results  of its application to electromagnetically coupled waves in a simplified liquid metal battery of rectangular shape. 
The results primarily concern the linear stages of one particular instability in a simplified situation when the base state has flat interfaces and zero flow, but the model itself has much broader applicability. It is well suited for analysis of more general situations: with background melt flows and interface deformations, spatially complex or even time-dependent  distributions of base electric current and magnetic field, etc.

The instability caused by the interaction between the externally generated vertical magnetic field and horizontal electric current perturbations associated with the interface deformations is demonstrated and analyzed. The growing perturbations have the form of rotating large-scale interfacial waves. The instability mechanism is similar to the mechanism of the rolling pad instability known for the aluminum reduction cells.   

In the case when the density jump at one interface is much smaller than at the other, only one interface is significantly deformed in the course of the perturbation growth, and the similarity to the behavior of reduction cells is quite close. In particular, the instability is controlled by the same non-dimensional groups: the horizontal aspect ratio (\ref{pargam}), the instability parameter (\ref{pi1}), and the Reynolds number (\ref{reynolds}). {Critically, the instability parameter (\ref{pi1}) is proportional to the square of the horizontal size of the cell.}

In the case when the density jumps at the two interfaces are comparable, both the interfaces are significantly deformed, and the behavior of the system is more complex and quite different from that of a reduction cell. The two interfacial waves can be coupled either symmetrically or antisymmetrically with a small phase shift between the waves. The presence of the second deformable interface may have a stabilizing effect. 

The model used in this study is idealized and cannot produce a complete and quantitatively accurate picture of a real system. Further work is advised, preferably in the way of experiments performed with large battery cells and three-dimensional numerical models based on accurate representation of the geometry and the complex distributions of the base current and magnetic field. Eigenvalue analysis of the simplified 2D and 3D system would also be interesting.

Even at this point, however, the available results can be utilized to make a preliminary assessment of the potential effect of the rolling pad instability on battery design. 
We have seen above that the unstable perturbations tend to grow strongly resulting in high amplitudes of interface deformation or even rupture of the electrolyte layer. The situation is dissimilar to that predicted for the thermal convection or Tayler instability in the sense that 
 reliable uninterrupted operation of a battery is only possible if the instability is avoided. Our conclusion is that the instability is a limiting factor for the versions of the battery design with small density difference between the metal and the electrolyte. One such version is the Mg-Sb battery \cite{Kim:2013}, in which $\rho^A\approx 1577$ kg/m$^3$,  $\rho^E\approx 1715$ kg/m$^3$, $\rho^B\approx 6270$ kg/m$^3$. Taking, as an example, the relatively stable geometry of a rectangular cell with $\Gamma=2.0$ and using the criterion $\Pi_{cr}=3.7$ derived in section \ref{sec:padsingle} we find that at typical $J_0=10^4$ A/m$^2$, $B_0=0.001$ T, a battery with $H_0^A=H_0^B=0.1$ m, $H_0^E=0.005$ m becomes unstable at the {horizontal size $L_x$ exceeding 0.7077 m.} This is larger than in the currently built laboratory prototypes or commercial concepts, but certainly in the range anticipated for future large-scale energy storage facilities \cite{Bradwell:2012}. 
 
 We should also mention that, similarly to the practice of the aluminum smelting industry, the instability can be avoided by rearranging the cell's wiring so as to reduce the vertical component of the magnetic field within the cell. Computational simulation tools built on the basis of our model can be utilized to predict the need for such a rearrangement and to assist in its optimization.

The instability appears less important for other battery concepts, in which $\Delta \rho^A$ is substantially larger. The results presented in section  \ref{sec:paddouble} suggest strong stabilization effect. For example, increasing $\Delta \rho^A$ in the example above to realistically possible 2000 kg/m$^3$ we find $L_x=3.5$ m, which does not appear to impose a serious limit on the technology.

\paragraph{Acknowledgements}
The author is thankful to Norbert Weber, Tom Weier, Valdis Bojarevics, Janis Priede, and Gerrit Horstmann  for interesting and stimulating discussions.
 Financial support was provided by the US NSF (Grant CBET 1435269).


\begin{thebibliography}{10}
\providecommand{\url}[1]{{#1}}
\providecommand{\urlprefix}{URL }
\expandafter\ifx\csname urlstyle\endcsname\relax
  \providecommand{\doi}[1]{DOI~\discretionary{}{}{}#1}\else
  \providecommand{\doi}{DOI~\discretionary{}{}{}\begingroup
  \urlstyle{rm}\Url}\fi

\bibitem{Mudpack}
Adams, J.C.: Multigrid software for elliptic partial differential equations:
  Mudpack ({1991}).
\newblock {NCAR Technical Note-357+STR}

\bibitem{Bojarevics:1994}
Bojarevics, V., Romerio, M.V.: {Long waves instability of liquid
  metal-electrolyte interface in aluminium electrolysis cells: a generalization
  of Sele's criterion}.
\newblock Eur. J. Mech. B, Fluids \textbf{13}(1), 33--56 (1994)

\bibitem{Bojarevics:2017}
Bojarevics, V., Tucs, A.: {MHD of Large Scale Liquid Metal Batteries}.
\newblock In: Light Metals 2017, pp. 687--692. Springer (2017)

\bibitem{Bradwell:2012}
Bradwell, D.J., Kim, H., Sirk, A.H.C., Sadoway, D.R.: Magnesium--antimony
  liquid metal battery for stationary energy storage.
\newblock Journal of the American Chemical Society \textbf{134}(4), 1895--1897
  (2012)

\bibitem{Davidson:2016}
Davidson, P.A.: Introduction to magnetohydrodynamics.
\newblock Cambridge University Press (2016)

\bibitem{Davidson:1998}
Davidson, P.A., Lindsay, R.I.: {Stability of interfacial waves in aluminium
  reduction cells}.
\newblock J. Fluid Mech. \textbf{362}, 273--295 (1998)

\bibitem{Herreman:2015}
Herreman, W., Nore, C., Cappanera, L., Guermond, J.L.: {Tayler instability in
  liquid metal columns and liquid metal batteries}.
\newblock J. Fluid Mech. \textbf{771}, 79--114 (2015)

\bibitem{Horstmann:2017}
Horstmann, G.M., Weber, N., T., W.: {Coupling and stability of interfacial
  waves in liquid metal batteries}.
\newblock Tech. rep. (2017)

\bibitem{Karniadakis:1991}
Karniadakis, G., Israeli, M., Orszag, S.: {High-order splitting methods for the
  incompressible Navier-Stokes equations}.
\newblock J. Comp. Phys. \textbf{97}(2), 414--443 (1991)

\bibitem{Kelley:2014}
Kelley, D.H., Sadoway, D.R.: {Mixing in a liquid metal electrode}.
\newblock Phys. Fluids \textbf{26}(5), 057,102 (2014)

\bibitem{Kim:2013}
Kim, H., Boysen, D.A., Newhouse, J.M., Spatocco, B.L., Chung, B., Burke, P.J.,
  Bradwell, D.J., Jiang, K., Tomaszowska, A.A., Wang, K., Wei, W., Ortiz, L.A.,
  Barriga, S.A., Poizeau, S.M., Sadoway, D.R.: Liquid metal batteries: Past,
  present, and future.
\newblock Chemical Reviews \textbf{113}(3), 2075--2099 (2013)

\bibitem{Kim:2013calcium}
Kim, H., Boysen, D.A., Ouchi, T., Sadoway, D.R.: Calcium--bismuth electrodes
  for large-scale energy storage (liquid metal batteries).
\newblock J. Power Sources \textbf{241}, 239--248 (2013)

\bibitem{Moreau:1984}
Moreau, R., Evans, J.W.: An analysis of the hydrodynamics of aluminum reduction
  cells.
\newblock J. Electrochem. Soc. \textbf{131}(10), 2251--2259 (1984)

\bibitem{Ouchi:2016}
Ouchi, T., Kim, H., Spatocco, B.L., Sadoway, D.R.: Calcium-based multi-element
  chemistry for grid-scale electrochemical energy storage.
\newblock Nature communications \textbf{7} (2016)

\bibitem{Rudiger:2007}
R\"{u}diger, G., Schultz, M., Shalybkov, D., Hollerbach, R.: {Theory of
  current-driven instability experiments in magnetic Taylor-Couette flows}.
\newblock Phys. Rev. E \textbf{76}, 056309 (2007)

\bibitem{Seilmayer:2012}
Seilmayer, M., Stefani, F., Gundrum, T., Weier, T., Gerbeth, G., Gellert, M.,
  R\"{u}diger, G.: {Experimental evidence for a transient Tayler instability in
  a cylindrical liquid-metal column}.
\newblock Phys. Rev. Lett. \textbf{108}, 108244501 (2012)

\bibitem{Sele:1977}
Sele, T.: {Instabilities of the metal surface in electrolyte alumina reduction
  cells}.
\newblock Met. Mat. Trans. B \textbf{8}, 613 (1977)

\bibitem{Shen:2016}
Shen, Y., Zikanov, O.: {Thermal convection in a liquid metal battery}.
\newblock Theor. Comp. Fluid Dyn. \textbf{30}(4), 275--294 (2016)

\bibitem{Sneyd:1994}
Sneyd, A., Wang, A.: {Interfacial instability due to MHD mode coupling in
  aluminium reduction cells}.
\newblock J. Fluid Mech. \textbf{263}, 343--360 (1994)

\bibitem{Stefani:2011}
Stefani, F., Weier, T., Gundrum, T., Gerbeth, G.: {How to circumvent the size
  limitation of liquid metal batteries due to the Tayler instability}.
\newblock Energy Conv. Manag. \textbf{52}(8-9), 2982--2986 (2011)

\bibitem{Sun:2004}
Sun, H., Zikanov, O., Ziegler, D.P.: {Non-linear two-dimensional model of melt
  flows and interface instability in aluminum reduction cells}.
\newblock Fluid Dyn. Res. \textbf{35}(4), 255--274 (2004)

\bibitem{Urata:1985}
Urata, N.: Magnetics and metal pad instability.
\newblock In: Essential Readings in Light Metals, pp. 330--335. Springer (2016)

\bibitem{Wang:2014}
Wang, K., Jiang, K., Chung, B., Ouchi, T., Burke, P.J., Boysen, D.A., Bradwell,
  D.J., Kim, H., Muecke, U., Sadoway, D.R.: {Lithium-antimony-lead liquid metal
  battery for grid-level energy storage.}
\newblock Nature \textbf{514}(7522), 348--50 (2014)

\bibitem{Weber:2016}
Weber, N., Beckstein, P., Herreman, W., Horstmann, G.M., Nore, C., Stefani, F.,
  Weier, T.: Sloshing instability and electrolyte layer rupture in liquid metal
  batteries.
\newblock Physics of Fluids \textbf{29}(5), 054,101 (2017)

\bibitem{Weber:2015}
Weber, N., Galindo, V., Priede, J., Stefani, F., Weier, T.: {The influence of
  current collectors on Tayler instability and electro-vortex flows in liquid
  metal batteries}.
\newblock Phys. Fluids \textbf{27}(1), 014103 (2015)

\bibitem{Weber:2014}
Weber, N., Galindo, V., Stefani, F., Weier, T.: {Current-driven flow
  instabilities in large-scale liquid metal batteries, and how to tame them}.
\newblock J. Power Sources \textbf{265}, 166--173 (2014)

\bibitem{Xiang:2017}
Xiang, L., Zikanov, O.: Subcritical convection in an internally heated layer.
\newblock Phys. Rev. Fluids \textbf{2}, 063,501 (2017)

\bibitem{Xu:2016}
Xu, J., Kjos, O.S., Osen, K.S., Martinez, A.M., Kongstein, O.E., Haarberg,
  G.M.: {Na-Zn liquid metal battery}.
\newblock J. Power Sources \textbf{332}, 274--280 (2016)

\bibitem{Zikanov:2015}
Zikanov, O.: Metal pad instabilities in liquid metal batteries.
\newblock Phys. Rev. E \textbf{92}(6), 063,021 (2015)

\bibitem{Zikanov:2000}
Zikanov, O., Thess, A., Davidson, P.A., Ziegler, D.P.: {A New Approach to
  Numerical Simulation of Melt Flows and Interface Instability in Hall –
  Heroult Cells}.
\newblock Met. Mat. Trans. B \textbf{31}, 1541--1550 (2000)

\end{thebibliography}

\end{document}